\documentclass[fleqn,usenatbib]{mnras}
\DeclareUnicodeCharacter{2212}{\textminus}

\usepackage{newtxtext,newtxmath}

\usepackage[T1]{fontenc}
\usepackage{booktabs}
\usepackage{multirow}
\usepackage[justification=centering]{caption}
\usepackage[para]{threeparttable}
\usepackage{amsmath}

\DeclareRobustCommand{\VAN}[3]{#2}
\let\VANthebibliography\thebibliography
\def\thebibliography{\DeclareRobustCommand{\VAN}[3]{##3}\VANthebibliography}

\usepackage{graphicx}	
\usepackage{amsmath}	

\title[Short title, max. 45 characters]{On the use of polarized thermal emission to constrain cloud grain size and temperature structure of sub-stellar objects} 

\author[Fei Wang]{
Fei Wang,$^{1,2,3}$ \thanks{E-mail:f.wang5@herts.ac.uk}
Yuka Fujii,$^{2}$
Ben Burningham$^{1}$
and Jinping He$^{3,4,5}$ \thanks{E-mail:jphe@niaot.ac.cn}
\\
$^{1}$ Department of Physics, Astronomy and Mathematics, University of Hertfordshire, AL10 9AB, Hatfield, United Kingdom.\\
$^{2}$ Division of Science, National Astronomical Observatory of Japan, 2-21-1 Osawa, Mitaka, Tokyo 181-8588, Japan\\
$^{3}$Laboratory of Solar and Space Instruments, Nanjing Institute of Astronomical Optics \& Technology, Chinese Academy of Sciences, Nanjing 210042, China \\
$^{4}$CAS Key Laboratory of Astronomical Optics \& Technology, Nanjing Institute of Astronomical Optics \& Technology, Chinese Academy of Sciences, Nanjing 210042, China \\
$^{5}$ University of Chinese Academy of Sciences, Nanjing 211135, China
}

\pubyear{2024}

\begin{document}
\label{firstpage}
\pagerange{\pageref{firstpage}--\pageref{lastpage}}
\maketitle

\begin{abstract}
Emission spectroscopy is an invaluable tool for probing the atmospheres of brown dwarfs and exoplanets, but interpretations based on flux spectra alone often suffer from degeneracies among temperature structure, chemical composition, and cloud properties. Thermal emission spectropolarimetry offers complementary sensitivity to these atmospheric characteristics. Previous studies have shown that linear polarization in fixed bandpasses depends on emission angle, temperature profile, and cloud scattering. In this study, we revisit these dependencies, emphasizing the wavelength-dependent effects that shape polarized spectra. We show that thermal polarization spectrum is primarily governed by: (1) a combination of temperature, temperature gradient, and wavelength; (2) cloud particle size; and (3) cloud optical thickness. Using the 3D Monte Carlo radiative transfer code ARTES, we simulate polarization spectra from a modeled 1D atmosphere. We find that, for a fixed cloud optical thickness, the polarization exhibits peaks at size parameters $x \sim 0.2$ and $x \sim 1$. However, the dependence on cloud optical thickness is more pronounced and tends to dominate the broadband polarization. We further show that much narrower polarization features in molecular absorption band, can in principle trace the local temperature gradient  at the photosphere of each wavelength. Future low-resolution (R $\sim$100) spectropolarimeter operating at 1–2 $\mu$m with sensitivities of $\leq 10^{-5}$ would be able to capture these polarization features, and may provide a new diagnostic for breaking degeneracies that commonly affect flux-only retrievals. This work represents an incremental step toward the challenging goal of jointly interpreting atmosphere from both intensity and polarization spectra.
\end{abstract}

\begin{keywords}
brown dwarfs – methods: analytical – polarization – radiative transfer – scattering
\end{keywords}



\section{Introduction}


Brown dwarfs serve as excellent laboratories for studying atmospheric processes with complexities akin to those of giant planets, while avoiding challenges posed by stellar irradiation and low signal-to-noise (S/N) ratios. Their atmospheres offer valuable insights into composition and thermal structure through distinct spectral signatures. Spectroscopic observations in recent years have significantly advanced our knowledge of their atmospheric compositions, particularly for major C- and O-bearing species \citep{zhang202113co,molliere2020retrieving,molliere2022interpreting}. Key properties such as metallicity, C/O ratios, and more recently, carbon isotope ratios have been used to constrain the formation pathways of brown dwarfs, shedding light on their formation locations and migration histories \citep{oberg2011effects, madhusudhan2014toward, madhusudhan2017atmospheric,zhang2021uniform}.

Two primary approaches are used to interpret brown dwarf spectra: finding the best-fit one among the grids of self-consistent models and a more agnostic atmospheric retrieval. The former is more traditional and involves interpolating pre-computed model grids to derive atmospheric and bulk properties (e.g., \citealt{marley1996atmospheric,burrows1997nongray,chabrier2000evolutionary,ackerman2001precipitating,allard2001limiting,baraffe2002evolutionary,burrows2003beyond,burrows2011dependence,morley2014water,zhang202113co,zhang2021uniform}). This method enables the inference of parameters such as surface gravity, mass, and effective temperature (e.g., \citealt{burgasser2006method,cushing2008atmospheric,kasper2009testing,king2010indi,bowler2010sdss,hinkley2015early,franson2023dynamical}). However, self-consistent model grids, despite incorporating sophisticated chemistry and physics, often systematically under- or overestimate the masses of brown dwarfs when compared to objects with measured dynamical masses \citep{dupuy2009dynamical,konopacky2010high,cheetham2018discovery,beatty2018significant,rickman2020spectral,bowler2021mcdonald,brandt2021improved}. Additionally, the fixed dimensions of grid models restrict the interpretation of object properties to predefined parameters. Remaining mismatches between observed data and model predictions across all available grid models suggest that some assumptions required by grid modeling may not be valid in many objects. 

In contrast, atmospheric retrieval offers a more flexible technique by directly constraining chemical abundances without assuming equilibrium chemistry. This allows for better insights into the metallicity and C/O ratio of brown dwarfs and directly imaged exoplanets. The ability to infer atmospheric properties in this way is crucial for understanding the formation, evolution, and chemical enrichment of substellar objects, much like what is seen in gas and ice giants within our solar system \citep{madhusudhan2016exoplanetary}.

The first comprehensive atmospheric retrieval study of two benchmark T dwarfs was conducted by \citet{line2015uniform}. Building on this pioneering effort, a series of atmospheric retrievals have since been performed on L, T, and Y dwarfs \citep{line2017uniform,burningham2017retrieval}, successfully constraining bulk properties across the T dwarf sequence \citep{lueber2022retrieval,zalesky2022uniform} and extending down to Y dwarfs \citep{zalesky2019uniform}. More recently, with the advent of data from the James Webb Space Telescope (JWST), atmospheric retrievals have been extended to the coldest Y dwarfs. \citet{kothari2024probing} conducted retrievals on the Y0 brown dwarf WISE J035934.06-540154.6 using JWST's low-resolution 0.96-12 $\rm \mu m$ spectrum. This broad-wavelength infrared spectrum provided 3-5 times more precise constraints on the mixing ratios of key gas-phase absorbers (H\(_2\)O, CH\(_4\), CO, CO\(_2\), PH\(_3\), and H\(_2\)S) compared to earlier studies that relied on HST data \citep{zalesky2019uniform}. Observations with JWST/NIRSpec of the coldest brown dwarf (WISE 0855−0714, 285 K) observed to date revealed the presence of methane, water vapor, ammonia, and carbon monoxide, suggesting evidence of disequilibrium chemistry in the atmosphere \citep{luhman2023jwst}. Furthermore, \citet{kuhnle2024water} detected water depletion and \(^{15}\)NH\(_3\) in the atmosphere of WISE 0855 using JWST/MIRI data. However, the observations did not detect any water ice clouds, and the spectrum is well matched by a cloudless model.

The above mentioned retrieval studies of ultra-cool brown dwarfs often involve a large number of free parameters and often suffer from degeneracies among atmospheric properties, particularly when addressing clouds and disequilibrium chemistry. The degeneracy could be more prominent when the data is restricted to narrow bandpasses \citep{wang2022unveiling}. A prominent degeneracy in brown dwarf flux retrievals arises between the temperature structure and gas abundances. Both the temperature profile and the abundances of dominant gases play a critical role in shaping the observed spectrum. The slope of the deepest part of the pressure-temperature ($P\text{--}T$ ) profile significantly influences the high-flux regions of the spectrum, which originate in the high-pressure layers of the atmosphere. Simultaneously, these flux peaks are shaped by the abundances of the most active gas species. Consequently, variations in the temperature structure can mimic changes in gas abundances, complicating efforts to disentangle the two effects based solely on thermal flux observations \citep{gandhi2018retrieval}. Furthermore, the presence of clouds exacerbate this degeneracy, as the $P\text{--}T$  profile can compensate for the presence of unmodeled clouds by reducing the temperature gradient in the photospheric region, thereby influencing the recovered gas abundance \citep{burrows2006and,ackerman2001precipitating}.

\citet{wang2020chemical,wang2022retrieving} presented atmospheric retrievals of HR 8799c, demonstrating that the retrieved atmospheric composition (e.g., O/H, C/O) differed significantly from previous studies, primarily due to different assumptions about the pressure-temperature profile. Similarly, \citet{phillips2024retrieving} conducted retrievals on two young, cloudy, red L dwarfs—CWISER J124332.12+600126.2 and WISEP J004701.06+680352.1. Their retrievals, which tested varying assumptions about the $P\text{--}T$  profile, were unable to consistently constrain the abundance measurements and C/O ratios, often yielding implausibly high abundances (e.g., H\(_2\)O).

Given this context, ensuring the accuracy and reliability of flux retrieval interpretations is critical for robust characterization of brown dwarf atmospheres. Polarimetry provides a promising and effective method to cross-validate interpretations and resolve degeneracies in flux retrievals, as it probes physical aspects of light not measured in nonpolarimetric photometry or flux spectroscopy. Radiation in the planetary atmosphere is scattered depending on the properties of the scattering particles. This type of scattering influences not only the flux but also the state of polarization. Unlike scalar intensity measurements, polarimetry quantifies light as a vector, based on the orientation of the electric field oscillations. This makes it highly sensitive to the physical and microphysical properties of an atmosphere (e.g., \citealt{hansen1974interpretation,hansen1974light}).

The diagnostic power of polarization in reflected light was first demonstrated by \citet{hansen1974interpretation}, who used it to determine the composition and size of cloud particles in Venus' atmosphere. The potential of this technique for detecting and characterizing exoplanets in reflected light has been widely recognized, with several studies providing numerical and analytical predictions (\citealt{seager2000photometric}, \citealt{stam2004using}, \citealt{buenzli2009grid}, \citealt{marley2011probing}, \citealt{karalidi2013flux}). \citet{lietzow2022polarimetric} performed a polarimetric investigation of 25 selected cloud compositions in exoplanetary atmospheres at optical to near-infrared wavelengths (0.3 to 1 $\rm \mu m$). Simulations show that most of these cloud condensates, such as chlorides, sulfides, or silicates can be distinguished from each other by their unique changes in net polarization as a function of wavelength, as well as a change in the sign of the polarization at specific wavelengths. These features could serve as important markers for characterizing cloud compositions in exoplanetary atmospheres.

On the other hand, brown dwarfs and self-luminous planets emit thermal radiation, which is scattered by atmospheric gases and cloud particles, polarizing the thermal radiation. Theoretical studies suggest that polarized thermal radiation in the near-infrared is generally more sensitive to spatial variability in a cloudy planet's atmosphere than thermal flux, potentially revealing important dynamical processes. \citet{de2011characterizing} studied the effects of individual atmospheric parameters, such as temperature gradients and cloud scattering properties, on the polarization of planetary thermal radiation at 1.05 and 1.11 $\rm \mu m$. They showed that the degree of linear polarization of an exoplanet’s thermal radiation is expected to be highest near the planet's limb. The degree of linear polarization depends on the temperature, its gradient, the scattering properties, and the distribution of cloud particles. Polarized thermal emission is typically stronger in molecular absorption bands \citep{stolker2017polarized}. The polarized thermal emission of brown dwarfs was successfully detected in 2020 using the Very Large Telescope/NaCo \citep{millar2020detection}. H-band linear polarization measurements of the nearby L/T transition binary Luhman 16 revealed degrees of polarization of \( p_A = 0.031\% \pm 0.004\% \) for Luhman 16A and \( p_B = 0.010\% \pm 0.004\% \) for Luhman 16B. These results suggest that Luhman 16A likely hosts longitudinal cloud structures.

However, most previous observational and theoretical investigations have focused on polarization within fixed photometric bands, neglecting the full spectral dependence of scattering processes. As a result, the influence of wavelength-dependent scattering properties on disk-resolved polarization spectra and how their combined effects may modulate resolved polarization feature remains unclear. In this study, we investigate how the polarized spectrum responds to the wavelength dependence of key atmospheric parameters, including the temperature structure, molecular opacity, and cloud scattering properties. We show how these parameters collectively influence both the broadband components of the polarization spectrum (those that vary gradually with wavelength) and the narrow spectral features. The resulting sensitivities offer potential constraints on cloud particle sizes and local temperature gradients, the latter providing information complementary to the $P\text{--}T$  profile typically inferred from intensity spectra but often subject to degeneracies with atmospheric composition.
From a detectability perspective, however, only the disk-integrated polarization spectrum is observable for unresolved brown dwarfs. Interpreting this net signal is challenging because it reflects the integrated contributions of regional disk-resolved polarization and large-scale atmospheric asymmetries. As an incremental step toward the broader goal of jointly retrieving atmospheric information from both intensity and polarization spectra, we assess the detectability of the predicted regional polarization features in the disk-integrated polarization spectrum.

Section~\ref{sec:2} examines key wavelength-dependent factors affecting spectropolarimetry in thermal emission. In Section~\ref{sec:3}, we describe the atmospheric parameterization, and the experimental setup to explore how the wavelength dependence of key atmospheric properties influences spectropolarimetry. Sections~\ref{sec:4} show the individual dependence of linear polarization on each key parameters. Section~\ref{sec:5} explores the combined effects of these factors on the resulting polarization spectrum and discusses their implications for atmospheric characterization, highlighting the potential of spectropolarimetry to preliminarily constrain cloud grain size and temperature structure. Finally, Section~\ref{sec:6} discusses our findings and evaluate the detectability of the predicted polarization features with current spectropolarimetry.

\section{Key Factors Affecting Thermal Emission Spectropolarimetry}\label{sec:2}

To investigate how wavelength-dependent processes shape polarization spectra, we begin by identifying the fundamental atmospheric factors that influence thermal emission spectropolarimetry. In substellar atmospheres, the polarization signal primarily arises from single scattering by gas molecules, clouds, and hazes. In the near- to mid-infrared regime relevant to thermal emission, polarization is mainly produced by single scattering from larger cloud particles. Multiple scattering generally acts to reduce the net polarization; Each additional scattering event tends to randomize the orientation of the polarization vector, leading to depolarization of the emergent light. In the following sections, we examine the key atmospheric parameters that control both the strength of polarization from single scattering and the degree of depolarization due to multiple scattering.

\begin{figure}
	\includegraphics[width=1.02\columnwidth]{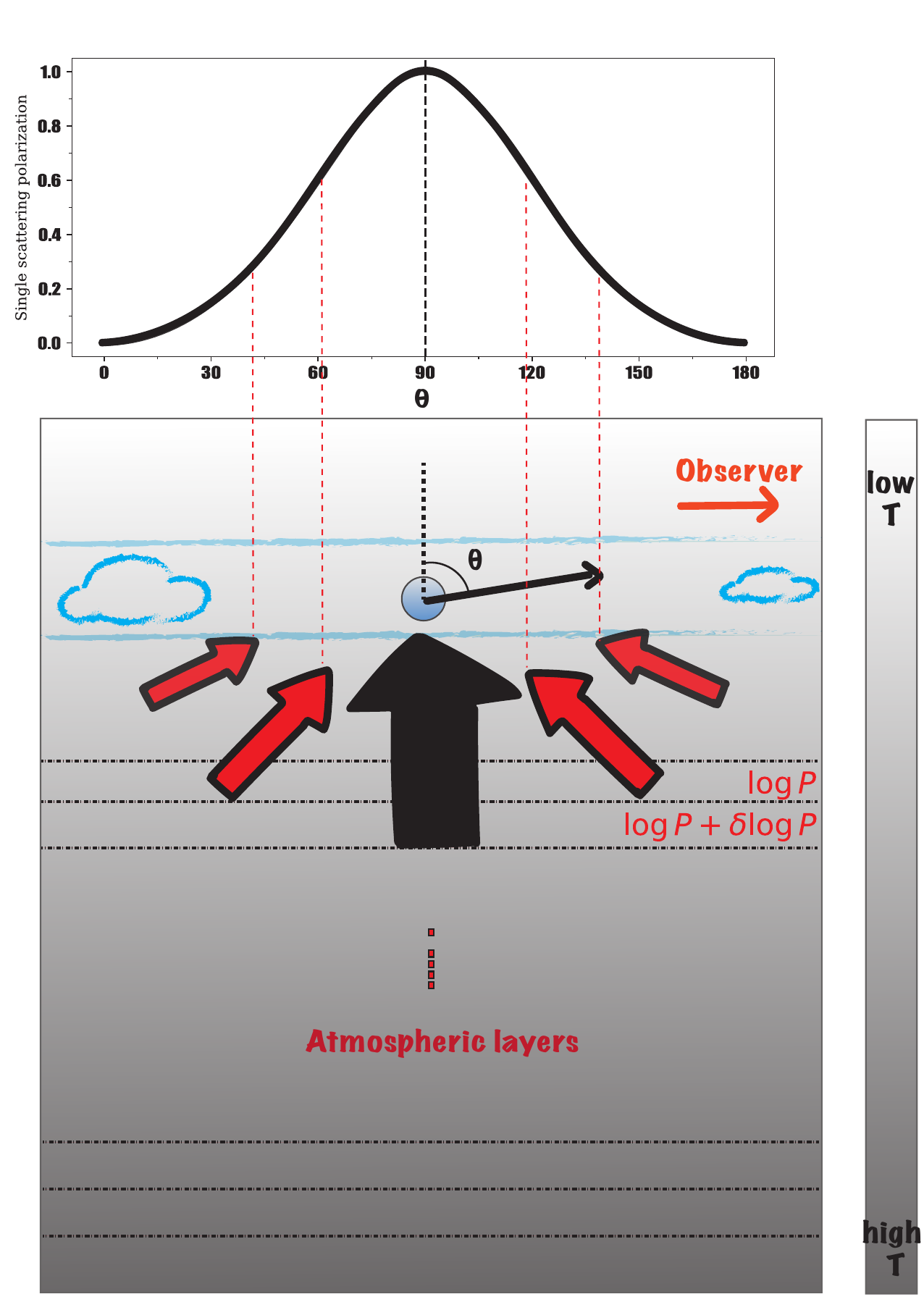}
    \caption{Schematic diagram of the thermal radiation process.  Photons emitted by thermal radiation from the lower atmosphere are scattered by cloud particle in the high atmosphere. The dark area represents regions of higher temperature and thus larger thermal radiation flux. Photons traveling perpendicular to the atmospheric layering (illustrated by the black arrow) originate from a lower altitude compared to photons traveling at other angles (illustrated by the red arrow). Upper panel gives the single scattering polarization of different scattering angles.}
    \label{fig1}
\end{figure}

\subsection{Parameters Affecting Single Scattering Polarization}\label{sec:2.1}
The polarization produced by single scattering in a substellar atmosphere depends sensitively on the scattering behavior of cloud particles, which is governed by the size parameter $x = 2 \pi r_{\rm eff} / \lambda$ , where \(\rm  r_{\rm eff} \) is the effective particle radius and \( \lambda \) is the wavelength of light. 
In the Rayleigh scattering regime where particles are much smaller than the wavelength (\( x \ll 1 \)), the scattering cross section exhibits strong wavelength dependence, with shorter wavelengths scattering more efficiently. As the particle size approaches or exceeds the wavelength (\( x \gtrsim 1 \)), scattering transitions into the Mie regime. Mie scattering is characterized by a weaker wavelength dependence and a strong forward-scattering peak, along with complex angular features such as rainbows or glories.

Rayleigh and Mie scattering produce distinct single-scattering phase functions and polarization characteristics. As illustrated in the upper panel of Figure \ref{fig1}, Rayleigh scattering yields a bell-shaped polarization efficiency curve, reaching 100\% at a scattering angle of \( 90^\circ \). In the context of thermal emission, this geometry is particularly relevant near the limb of the disk, where photons emitted from deeper atmospheric layers are scattered by high-altitude cloud particles at angles close to \( 90^\circ \) with respect to the observer. This configuration produces a strong local polarization signal, making the planetary limb the region of peak polarization intensity.

To further illustrate this, Figure \ref{fig1} includes a schematic diagram of the atmosphere and single scattering geometry in the limb region. The atmosphere is modeled as a set of plane-parallel layers with an additional high-altitude cloud layer overlying them. Thermal radiation is emitted in various directions from different depths in the atmosphere. For a given optical depth (e.g., \( \tau = 1 \)), photons traveling vertically upward originate from deeper layers (black arrow), whereas photons emitted at oblique angles reach the cloud layer from shallower altitudes (red arrows). These differences in emission origin influence the scattering angle relative to the observer and, in turn, the resulting polarization strength. Photons emitted vertically from deeper layers and scattered by cloud particles at near-\( 90^\circ \) angles generate strong polarization signals. In contrast, photons emitted obliquely from higher altitudes are scattered across a wider range of angles, typically yielding weaker polarization. Therefore, the ratio of vertically emitted radiance to oblique radiance, denoted as \( I_{\rm ratio} \), serves as a key diagnostic of the overall polarization strength from single scattering.

To clarify how the \( I_{\rm ratio} \) relates to the atmospheric structure, we derive an analytical expression based on the thermal emission from adjacent layers in the atmosphere.  
Consider two neighboring sub-layers as illustrated in Figure~\ref{fig1}: an upper layer at pressure \(\log P\), where photons are emitted at oblique angles, and a deeper layer at a slightly higher pressure \((\log P + \delta \log P)\), from which photons travel vertically upward toward the cloud layer.

According to Planck’s law, the spectral radiance \(I(\lambda, T)\) is given by
\begin{equation}
I(\lambda,T) = \frac{2hc^2}{\lambda^5}
\frac{1}{\exp\!\big(hc/(\lambda k T)\bigr) - 1}.
\label{equa1}
\end{equation}

In the Wien limit \((hc/(\lambda k T) \gg 1)\), the intensity gradient with respect to logarithmic pressure can be expressed as
\begin{equation}
\begin{aligned}
\frac{dI(\lambda,T)}{d\log P}
&= \frac{dI(\lambda,T)}{dT} \, \frac{dT}{d\log P} \\[6pt]
&= \frac{2hc^2}{\lambda^5} \,
   \frac{hc}{\lambda k} \,
   \frac{1}{T^2} \,
   \exp\!\Big(-\frac{hc}{\lambda k T}\Big) \,
   \frac{dT}{d\log P} \\[6pt]
&= I(\lambda,T) \,
   \frac{hc}{\lambda k} \,
   \frac{1}{T^2} \,
   \frac{dT}{d\log P}.
\end{aligned}
\label{equa2}
\end{equation}

The intensity ratio between the two layers can be approximated by

\begin{equation}
\begin{aligned}
\frac{1}{I(\log P)} \frac{dI}{d\log P}
&\simeq
\frac{1}{I(\log P)} 
\frac{I(\log P + \delta \log P) - I(\log P)}{\delta \log P} \\[6pt]
&\simeq
\frac{\frac{I(\log P + \delta \log P)}{I(\log P)} - 1}{\delta \log P}.
\end{aligned}
\label{equa3}
\end{equation}
where \(\delta \log P\) is the pressure difference between the two layers.

Substituting Equation~\ref{equa2} into Equation~\ref{equa3} yields
\begin{equation}
I_{\rm ratio}=\frac{I(\log P + \delta \log P)}{I(\log P)}
\sim
\frac{hc}{\lambda k} \,
\frac{1}{T^2} \,
\frac{dT}{d\log P} \,
\delta \log P
\label{equa4}
\end{equation}

This relation highlights two key dependencies. First, the radiance ratio varies with wavelength. Second, it depends on the product (\(\frac{1}{T^{2}}\frac{dT}{d\log P} \)), which encodes both the local temperature and the vertical steepness of the temperature profile. Together with the cloud size parameter, these quantities jointly determine the strength of the single scattering polarization signal by shaping both the thermal emission field and the scattering behavior.

\subsection{Parameters Affecting the Extent of Depolarization in Multiple Scattering}\label{sec:2.2}

The extent to which multiple scattering depolarizes thermal emission is primarily governed by two key atmospheric parameters: the cloud optical depth, \( \tau_{\rm cld} \) and the single scattering albedo, \( \omega \). As photons propagate through a cloudy atmosphere, each scattering event can alter both their direction and polarization state. With successive scatterings, the polarization vectors become increasingly mixed, leading to the depolarization effect that reduce 
net degree of polarization.

The optical depth \( \tau_{\rm cld} \) quantifies the total extinction (scattering plus absorption) through the cloud layer. In clouds with low optical depth, photons are likely to escape after zero or one scattering, allowing single scattering to dominate. In contrast, high optical depth increases the likelihood of multiple interactions before a photon escapes, making multiple scattering more significant. The single scattering albedo \( \omega \) describes the probability that a photon is scattered rather than absorbed upon encountering a cloud particle. When \( \omega \) is close to unity and the optical depth is high, photons can undergo many scattering events, enhancing the depolarization effect.

Multiple scattering becomes important under conditions of both high optical depth and high albedo. To maximize the observable single scattering polarization while minimizing depolarization, there exists an optimal range of \( \tau_{\rm cld} \) and \( \omega \). The interplay between scattering and absorption, combined with the vertical distribution of cloud particles, plays a critical role in shaping the net polarization signal from a substellar atmosphere.

Table~\ref{tab:1} lists the key parameters of cloud particle size parameter, ratio of vertically emitted radiance to oblique radiance, and cloud optical thickness that influence the spectropolarimetry.

\begin{table*}
  \centering
  \begin{threeparttable}
    \centering
    \caption{Explored parameters that influence spectropolarimetry}
    \label{tab:1}
    \begin{tabular}{@{}ll@{}}
      \toprule
      \toprule
      Parameter & Sub-parameter \\ \midrule
     \multicolumn{1}{|l|}{\multirow{2}{*}{Cloud particle size parameter ($x$)}} & \multicolumn{1}{l|}{Single scattering polarization ($P_{\rm single}$)} \\ \cmidrule(l){2-2} 
      \multicolumn{1}{|l|}{} & \multicolumn{1}{l|}{Single scattering albedo ($\rm \omega$)} \\ \midrule

      \hspace*{0.15cm} Cloud optical thickness ($\rm \tau_{cld}$) & Cloud cross section\tnote{a}, Cloud mass density\tnote{b} \\ \midrule
      
      \multicolumn{1}{|c|}{\multirow{2}{*}{Ratio of vertically emitted radiance to oblique radiance (\(I_{\rm ratio}\))}} & \multicolumn{1}{l|}{Wavelength (\(\rm \frac{1}{\lambda}\))} \\ \cmidrule(l){2-2} 
      \multicolumn{1}{|c|}{} & \multicolumn{1}{l|}{  Temperature and its gradient (\(\frac{1}{T^{2}}\frac{dT}{d\log P} \))} \\    
\bottomrule
    \end{tabular}
    \begin{tablenotes}
      \item [a] Cloud cross section also depends on the cloud size parameter.
      \item [b] Total mass of cloud particles per unit volume, affects final cloud optical thickness.
    \end{tablenotes}
  \end{threeparttable}
\end{table*}

\section{Methods}\label{sec:3}

In the preceding section, we identified the key parameters that shape polarized spectra. Here, our goal is to investigate how the key atmospheric properties influence the wavelength dependence of the thermal emission polarization. To achieve this, we start from setting up a simple 1D plane-parallel atmosphere with parametrized T-P profile, cloud scattering properties and cloud vertical distribution. Then, using this plane-parallel atmosphere as an input, we employ a three-dimensional Monte Carlo radiative transfer code ARTES \citep{stolker2017polarized} to compute the resultant polarization spectrum.

Below, we describe in detail the ARTES code, the parameterized atmospheric setup for brown dwarf, and the experimental design used to explore the wavelength dependence of key atmospheric parameters.

\subsection{Monte Carlo Radiative Transfer Code: ARTES}\label{sec:3.1}

A complete description of ARTES is provided in \citet{stolker2017polarized}. Here, we briefly summarize its key features relevant to this study. ARTES requires a parameterized atmospheric structure as input, including the temperature–pressure profile, molecular opacities, and cloud scattering properties. To enable flexible control over these parameters, we introduced several modifications to ARTES. In particular, we implemented a parameterized $P\text{--}T$  structure that allows direct manipulation of the atmospheric temperature gradient. For gas opacities, besides the provided pre-tabulated grid for interpolating temperature-dependent molecular abundances and absorption cross sections, we also integrate the opacity database from PICASO \citep{batalha2019exoplanet, mukherjee2023picaso} to calculate the free-chemistry cross sections. The latter was used when generating the polarization spectrum for the free-retrieved gases abundances. 

ARTES models the radiation field using photon packets emitted from a spatially extended source. These photons propagate through a three-dimensional spherical grid, interacting with atmospheric particles via scattering and absorption, until they either exit the atmosphere or are absorbed. To efficiently simulate photon trajectories, ARTES employs the peel-off technique, projecting each scattering event onto the detector and weighting it according to the local phase function and optical depth \citep{yusef1984bipolar}.

To balance computational efficiency with sufficient signal-to-noise, we simulate $10^{9}$ photon packets per model. In the limb region of the disk, the S/N of the simulated polarization at 1.5 $\mu$m ranges between 20 and 50. Although the Monte Carlo method offers substantial flexibility for handling the complex geometries and physical conditions of substellar atmospheres, it faces limitations in very high optical depth regimes (\(\tau \gtrsim 100\)). In such cases, even with the modified random walk approach, photon noise increases and fluxes may be significantly underestimated \citep{krieger2024improving}. To avoid these issues, we constrain our study to cloud optical depths below 50.

\begin{figure*}
	\includegraphics[width=2.12\columnwidth]{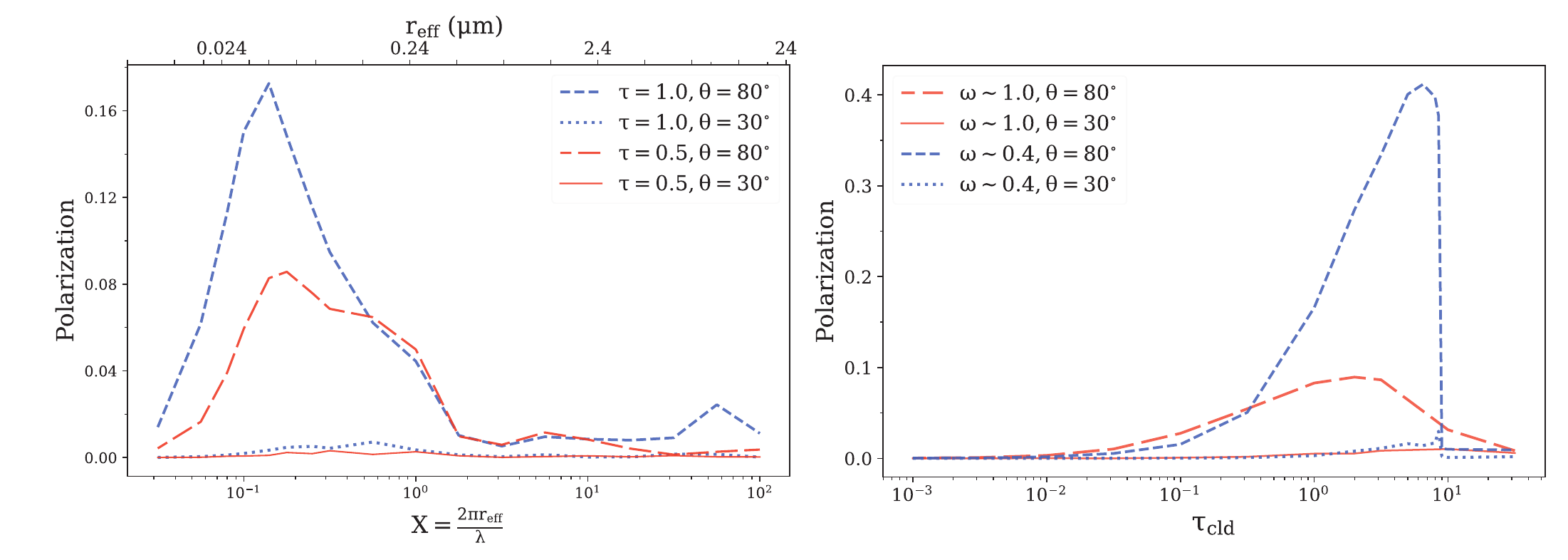}
    \caption{Polarization at $\lambda = 1.5 \, \mu \mathrm{m}$ is shown for atmospheres with a high-altitude cloud, illustrating the effects of (left) varying cloud grain size parameters and (right) different cloud optical thicknesses. Dashed lines correspond to emission angles of $80^\circ$, while dotted lines indicate $30^\circ$. In the left panel, two cloud optical thickness values are tested to examine their impact on the location of the polarization peak. Right panel suggests clouds with higher absorption (lower scattering albedo, $\omega$) can sustain significant polarization even at higher optical thicknesses ($\tau \ge 10$).
}
    \label{fig2}
\end{figure*}

\subsection{Atmospheric Parameterization}\label{sec:3.2}

In brown dwarf atmospheres, various opacity sources, both continuum and molecular, govern the absorption and scattering of thermal radiation. These opacities are strongly wavelength-dependent, causing different atmospheric layers to be probed across the spectrum. In spectral regions dominated by gas absorption, the higher opacity shifts the photosphere to higher altitudes compared to the continuum, where the atmosphere is more transparent. Our study focuses on the atmosphere of cloudy brown dwarfs, with a parameterization framework consisting of two main components: the thermal structure and the cloud properties. For the temperature profile, we adopt the parameterized prescription from \citet{de2011characterizing}, which enables controlled exploration of the impact of the temperature gradient on the polarization spectrum.

The atmosphere is divided into 50 layers evenly spaced in logarithmic pressure, spanning from \(10^{-4}\) bar at the top to \(10^2\) bar at the bottom. To allow comparisons across different atmospheric models, we assume simplified, ad hoc profiles under the assumption of hydrostatic equilibrium. Specifically, the temperature is held constant at pressures \( P < 10^{-3}\) bar and \( P > 1\) bar. In the intermediate pressure range from \(10^{-3}\) to 1 bar, where near-infrared emission primarily originates, we impose a constant temperature gradient, \(\frac{dT}{d\log P} \), centered at a pivot point of \(T = 1800\) K at \(P = 0.1\) bar. The temperature in each layer within this region is given by:

\begin{equation}
T_i = \frac{dT}{d \log P} \log p_i + C,
\label{eq5}
\end{equation}
where \(C\) is a normalization constant and \(dT/d\log P\) represents the temperature gradient. This formulation allows straightforward control over the vertical thermal structure.

To efficiently scatter near-infrared radiation and produce detectable polarization, sub-micron-sized particles are required. These particles operate in the Rayleigh regime (\(2\pi r_{\rm eff} \leq \lambda\)), where polarization efficiency is enhanced. We assume a cloud composition of enstatite (MgSiO\(_3\)), consistent with \citet{stolker2017polarized}, using complex refractive indices from \citet{dorschner1995steps}. Following Equation~\ref{eq5}, the baseline temperature profile is defined with a temperature gradient of 300~K per decade of pressure and a temperature of 1800~K at $P$ = 0.1 bar. The intersection between the condensation curve of MgSiO\(_3\) cloud and the baseline profile occurs near 1 bar ( $T\sim$ 1600~K), indicating that enstatite clouds could form above this pressure level. Forward-modeling and retrieval analyses of mid-infrared spectra of some L dwarfs have shown that reproducing the silicate feature at 8–10 $\mu$m may require silicate clouds to be lofted up to 10$^{-2}$ bar \citep{luna2021empirically,suarez2023ultracool}. Small silicate particles ($\leq 1 \mu$m) could stay lofted high in the atmosphere (lofting timescale shorter than the falling timescale) with vigorous enough mixing. However, the detailed vertical distribution of the cloud is governed by complex microphysical processes, including nucleation, sedimentation, and vertical mixing.

For simplicity, and to maximize single-scattering polarization within the atmosphere, we adopt a single high-altitude cloud layer located at located at \(P = 10^{-3}\) bar. Cloud particles are assumed to follow a gamma size distribution with a baseline effective radius \(r_{\rm eff} = 0.1\ \mu\mathrm{m}\) and effective variance \(v_{\rm eff} = 0.05\). The scattering and absorption properties are computed using Mie theory, assuming homogeneous spherical particles. The baseline wavelength is set to \(1.5\ \mu\mathrm{m}\), and the cloud optical depth is normalized to \(\tau = 1\) at this wavelength.

\subsection{Experimental Setup}\label{sec:3.3}

As discussed in Section ~\ref{sec:2}, key atmospheric parameters including the temperature gradient, cloud particle size parameter, cloud optical thickness, and single scattering albedo may significantly influence thermal spectropolarimetry. In this section, we detail the experimental setup used to explore how each of these factors affects the polarization of thermal emission. In Section ~\ref{sec:2.2}, we discussed that the single scattering albedo (\(\omega\)) plays a central role in depolarization through multiple scattering processes, thereby affecting the degree of polarization. Notably, \(\omega\) itself depends on the cloud particle size parameter (\(x\)) and the complex refractive index of the cloud particles. \citet{hansen1974light} established that the imaginary part of the refractive index (\(n_i\)) governs the absorption efficiency of cloud particles: higher \(n_i\) values lead to stronger absorption, resulting in smaller \(\omega\). 

In this study, we focus on clouds composed of MgSiO\(_3\), for which the complex refractive index is well characterized. When performing Mie scattering calculations for MgSiO\(_3\) clouds across different particle size distributions, \(\omega\) is primarily controlled by the parameter \(x\). Therefore, in our subsequent analysis, we use \(x\) as a proxy to explore how \(\omega\) influences multiple scattering and ultimately the resulting thermal polarization.

To isolate and examine the influence of each parameter, we analyze the linear polarization spectrum at the limb of the planetary disk, where thermal polarization is expected to be strongest. Specifically, we focus on polarization at scattering angles of $\theta = 80^\circ$. Our analysis adopts a baseline atmospheric temperature profile following the thermal structure parametrization outlined in Section ~\ref{sec:3.2}, with a temperature gradient of 300 and a temperature of 1800 K at $ P = 0.1$ bar. For comparison, the dry adiabatic lapse rate is $\nabla_{\text{ad}} =  d\log T / d\log P = 0.286$, corresponding to $ dT / d\log P \approx 428$ K at a temperature of 1500 K.

Based on this baseline temperature profile, the atmospheric opacity is interpolated from default chemical equilibrium grid that include the important absorbing species species like $\rm H_{2}O$, $\rm CO$, $\rm CO_{2}$, $\rm CH_{4}$, $\rm NH_{3}$, $\rm CrH$, $\rm FeH$, $\rm SiO$, $\rm Na$, $\rm K$. The continuum opacities include H\(_2\)-H\(_2\) and H\(_2\)-He collision-induced absorption, while Rayleigh scattering contributions from H\(_2\), He, and CH\(_4\) are also considered. In each experiment, we vary only one parameter while holding all others constant, as outlined below:

1. Dependence on cloud particle size parameter (\(x\)). 

We adopt the default temperature profile and fix the cloud optical thickness at 1 at 1.5 \(\mu \mathrm{m}\). The cloud particle size parameter \(x\) is varied from \(10^{-2}\) to \(10^{2}\), by adjusting the effective radius of the cloud particles accordingly.

2. Dependence on cloud optical thickness (\(\tau_{\rm cld}\)). 

\begin{figure*}
	\includegraphics[width=2.1\columnwidth]{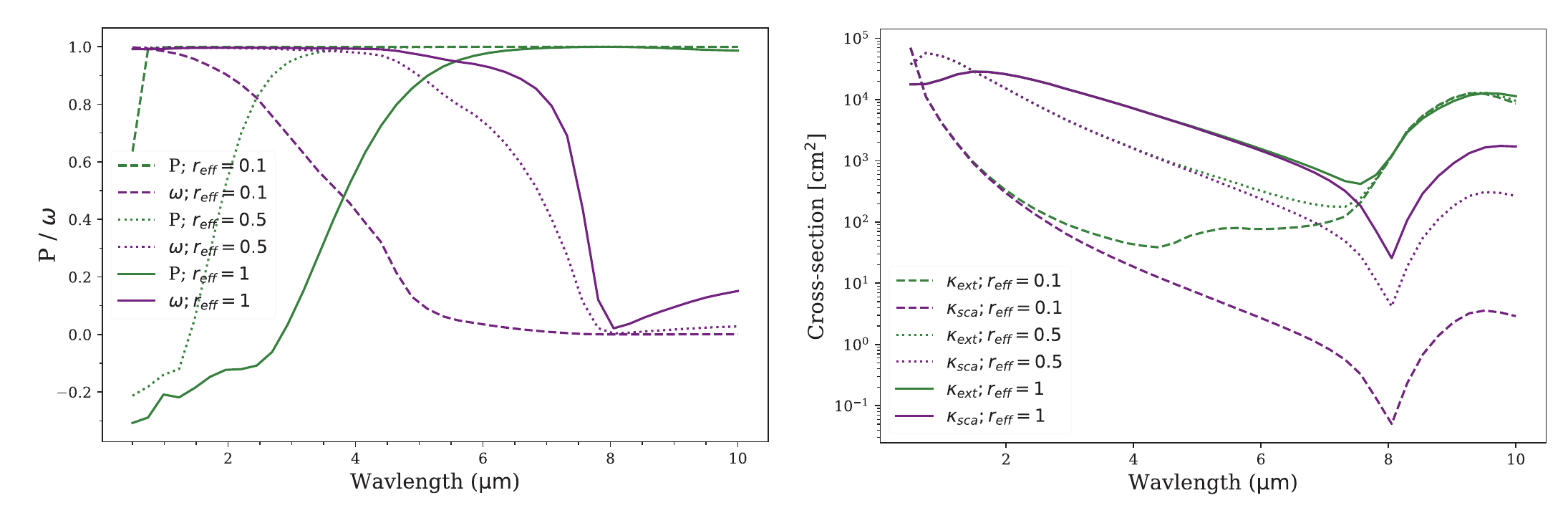}
    \caption{Left: spectral dependence of the single scattering polarization and single scattering albedo of $\rm MgSiO_{3}$ cloud in a gamma distribution of effective particle size of 0.1, 0.5 and 1 $\rm \mu m$ Right: spectral dependence of cloud extinction cross section ($\rm \kappa_{ext}$) and cloud scattering cross section ($\rm \kappa_{sca}$) for different grain sizes.}
    \label{fig3}
\end{figure*}

Basically, the cloud optical thickness can be determined by the cloud cross section and column density of the cloud particle. While the cloud cross section depends on the cloud size parameter, which is in general a function of wavelength. Thus, $\tau _{\rm cld}$ is wavelength-dependent. To explore how it affect the spectropolarimetry, we we assume both artificially constant $\tau _{\rm cld}$ and more realistic wavelength-dependent $\tau _{\rm cld}$ in this study. In the cases of wavelength-dependent $\tau _{\rm cld}$, we use the baseline temperature profile and fix cloud effective radius at 0.1 \(\mu \mathrm{m}\). The cloud optical thickness \(\tau_{\rm cld}\) varies from \(10^{-3}\) to 50, again at wavelength of 1.5 \(\mu \mathrm{m}\). The exact wavelength dependence of $\tau_{\rm cld}$ of MgSiO\(_3\) cloud is shown by the green dashed curve in the right panel of Figure~\ref{fig3}.

3. Dependence on wavelength and temperature gradient.

To explore the combined influence of wavelength and atmospheric thermal structure, we vary the parameters \( \frac{1}{\lambda} \frac{1}{T^{2}}\) and the temperature gradient \(\frac{dT}{d\log P} \). First, we generate a range of temperature-pressure profiles with temperature gradient varying from 50 to 300 using the parametrization from Section ~\ref{sec:3.1}. For each profile, we locate the atmospheric layer where the optical depth \(\tau = 1\), and record the corresponding temperature \(T_{\tau = 1}\). This temperature, combined with the desired value of \(\frac{1}{\lambda} \frac{1}{T^{2}}\), determines the wavelength \(\lambda\) under consideration. To isolate polarization effects, we adjust the cloud effective radius such that the particle size parameter remains fixed at \(x \approx 0.4\), and simultaneously tune the cloud mass density to maintain a constant \(\tau_{\mathrm{cld}} = 1\) at the corresponding wavelength.

4. Effect of molecular absorption.

Wavelength-dependent molecular absorption introduces additional structure in the polarization spectrum by altering the atmospheric opacity. This includes both continuous absorption and selective absorption by specific molecules. To evaluate the impact of molecular absorption, we compare two scenarios: (1) only continuum absorption and (2) continuum absorption combined with molecular absorption from water vapor. In both cases, the cloud effective radius fixed at 0.1 \(\mu \mathrm{m}\) and the cloud optical thickness set to 1 across the wavelength range of 1–2 \(\mu \mathrm{m}\). 

\section{Results}\label{sec:4}

This section presents how individual atmospheric parameters of cloud grain size parameter, cloud optical thickness, temperature structure and molecular absorption affect the linear polarization of thermal emission.

\subsection{Dependence on Cloud Particle Size Parameter}\label{sec:4.1}

The left panel of Figure \ref{fig2} shows how the degree of linear polarization ($P$) varies with the cloud particle size parameter across different regions of a planetary disk. Scattering angles of $\theta = 80^\circ$ and $\theta = 30^\circ$ correspond to the limb and central regions of the disk, respectively. At the center, the polarization signal remains near zero for all values of $x$, due to the low efficiency of single-scattering polarization in this region \citep{de2011characterizing}. In contrast, at the limb, the polarization more clearly reveals the influence of cloud particle size on thermal emission. As shown by the blue dashed curve in the left panel of Figure \ref{fig2}, the polarization degree peaks around $x \sim 0.2$. This peak results from an optimal single scattering albedo that enhances absorption, suppresses multiple scattering, and reduces depolarization, thereby maximizing the net polarization.

Specifically, the influence of cloud particle size on the polarization spectrum can be decomposed into two key components: single scattering polarization and the wavelength-dependent single-scattering albedo. Figure \ref{fig3} illustrates these components by showing the spectral dependence of single scattering polarization, single-scattering albedo, and extinction cross-section for $\rm MgSiO_3$ cloud grains with effective radii of 0.1, 0.5, and 1 $\rm \mu m$. At shorter wavelengths ($\lambda \leq 2\pi r_{\rm eff}$), Mie scattering dominates. In this regime, single scattering polarization increases monotonically with wavelength until the size parameter $x = 2\pi r_{\rm eff}/\lambda$ approaches unity. Beyond this point, polarization decreases rapidly with increasing wavelength. This behavior is evident in the left panel of Figure \ref{fig3}, where the single scattering polarization peaks near 0.62, 3.14, and 6.28 $\rm \mu m$ for particles of 0.1, 0.5, and 1 $\rm \mu m$, respectively.

The single-scattering albedo $\omega$ also varies with $x$. Generally, as $x$ increases in the Mie regime, the scattering cross-section increases nonlinearly (see Figure 20 in \citealt{lacy2020jwst}). For large particles ($x > 1$), $\omega$ approaches unity, indicating efficient scattering and low absorption (see left panel of Figure~\ref{fig3}). This promotes multiple scattering, which tends to depolarize the emergent radiation. Conversely, for small $x$, absorption dominates, reducing $\omega$. Around $x \sim 0.2$, $\omega$ drops to approximately 0.4. This enhanced absorption suppresses multiple scattering, reducing
depolarization effects and thereby increasing the net polarization.

\begin{figure*}
    \includegraphics[width=2.12\columnwidth]{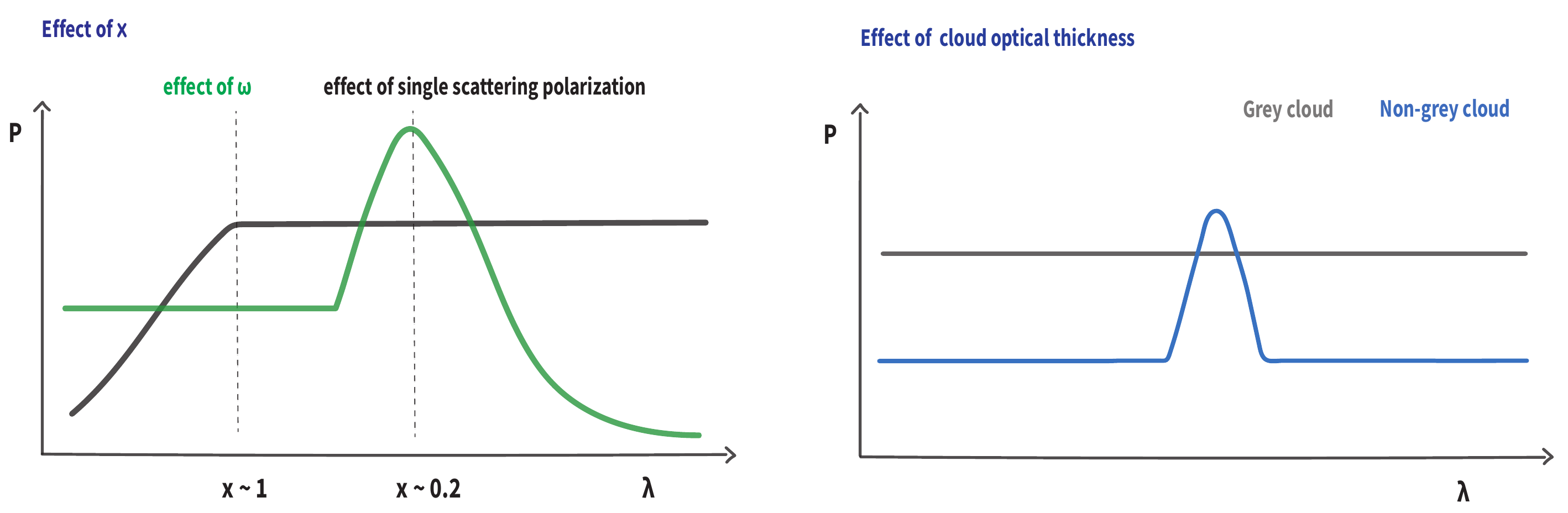}
    \caption{Schematic figure illustrates how cloud size parameter and cloud optical thickness influence the thermal polarization spectrum. The cloud size parameter determines the single scattering polarization and scattering albedo. Left panel shows the response of polarization to the changes to single scattering polarization and scattering albedo. Right panel: response of polarization to the changes to the cloud optical thickness. The artificially constant $\rm \tau_{cld}$ and more realistic wavelength dependent $\rm \tau_{cld}$ determined by the cloud cross section and column density of the cloud particle are shown with gray and blue curve respectively. For a non-gray cloud with a fixed column density, the wavelength dependence of $\tau_{\rm cld}$ is governed by the cloud’s refractive index and particle size distribution. Consequently, the polarization varies with wavelength, and the peak occurs at wavelengths where $\tau_{\rm cld}(\lambda)$ is in the regime that favors efficient single-scattering while suppressing multiple-scattering events.
}
    \label{fig4}
\end{figure*}

These effects are summarized in the left panel of Figure \ref{fig4}, where the contributions from single scattering polarization and albedo are shown as black and green curves, respectively. When each parameter is considered independently, two distinct polarization peaks emerge: one near $x \sim 1$, driven by the single scattering polarization, and another near $x \sim 0.2$, driven by enhanced absorption and reduced multiple scattering. Moreover, the product $ P_{\rm single} \cdot \omega$ quantifies the grain size most effective at contributing to the observed polarization at a given wavelength, since the Stokes parameters $Q$ and $U$ are proportional to this product in the single-scattering limit \citep{kataoka2015millimeter}. Thus, the interplay between $ P_{\rm single}$ and $\omega$ effectively encodes cloud particle size information, producing a characteristic broadband polarization feature. This broad band polarization signature provides a promising diagnostic of cloud grain sizes, which we explore further in Section~\ref{sec:5}.

Returning to the left panel of Figure \ref{fig2}, the polarization peak around $x \sim 0.2$ corresponds to the effect of  $\omega$ with enhanced absorption suppressing multiple scattering, as shown by the green curve in Figure \ref{fig4}. The red curves, which represent a reduced cloud optical thickness ($\tau = 0.5$), show a lower overall polarization due to fewer scattering events. While the peak position remains close to $x \sim 0.2$, it shifts slightly, suggesting that although the dominant factor setting the peak is the scattering/absorption transition, the efficiency of single scattering influenced by optical depth also affects the peak location. We examine this dependence in more detail in Section~\ref{sec:5.2}.

\subsection{Dependence on Cloud Optical Thickness}\label{sec:4.2}

The right panel of Figure \ref{fig2} shows how the degree of linear polarization responds to variations in cloud optical thickness (\(\tau_{\rm cld}\)). Initially, as \(\tau_{\rm cld}\) increases, more photons are scattered by cloud particles, which enhances the polarization signal. However, beyond a certain threshold, multiple scattering becomes increasingly dominant, leading to depolarization and a subsequent decline in \(P\). This results in a characteristic peak in the polarization curve. At very high optical depths, thermal photons originating from deeper atmospheric layers are absorbed and re-emitted by the overlying optically thick cloud. The re-emitted photons undergo extensive multiple scattering within the cloud, which significantly diminishes their polarization. Consequently, the polarization degree plateaus at low values, exhibiting minimal variation with further increases in optical depth.

To further explore this behavior, we also examine the influence of the cloud single scattering albedo on the polarization response caused by cloud optical depth. While the general trend remains consistent across different albedo values, the position of the polarization peak shifts depending on \(\omega\). For instance, in the case of a lower albedo (\(\omega = 0.4\)), enhanced absorption by cloud particles suppresses multiple scattering. This reduces the depolarization effect, allowing the polarization degree to remain relatively high even at larger optical depths (e.g., \(\tau \sim 10\)), as seen in the blue dashed curve in the right panel of Figure \ref{fig2}. For optically thin clouds, the impact on the emergent thermal flux is minimal, as these layers are nearly transparent to outgoing radiation. As a result, only a small fraction of photons are scattered by cloud particles, and the corresponding polarization is weak. As optical depth increases, two competing effects shape the polarization signal: absorption enhances \(P\) by reducing the intensity of unpolarized thermal emission, while increased multiple scattering suppresses \(P\) by depolarizing the emergent radiation. At very large optical depths (\(\tau > 10\)), the latter dominates, leading to the observed decline in polarization.

It is now evident that an optimal cloud optical thickness at a given wavelength can lead to a peak in polarization, provided the cloud particle size parameter allows for efficient single scattering. The cloud optical thickness, $\tau_{\rm cld}$, generally varies with wavelength, as it depends on the size parameter, as illustrated in the right panel of Figure \ref{fig3}. For $\rm MgSiO_3$, the extinction cross-section decreases with increasing wavelength, resulting in a negative slope. This indicates that the polarization peak associated with $\tau_{\rm cld}$ can shift depending on the cloud composition. Non-gray cloud species, shown by the blue curves in the right panel of Figure \ref{fig4}, exhibit distinct wavelength-dependent extinction gradients that can influence the location of the polarization peak. In contrast, gray clouds, which have a wavelength-independent $\tau_{\rm cld}$, yield a more uniform polarization signal across the spectrum.

\subsection{Dependence on Wavelength and Temperature Gradient}\label{sec:4.3}

Wavelength and atmospheric temperature gradient together govern the anisotropy of the radiation field reaching the cloud layer. Specifically, the ratio of vertically emitted radiance to oblique radiance. A higher ratio favors single scattering events that occur at scattering angles near \(90^\circ\), thereby enhancing the degree of linear polarization. However, these two parameters are intrinsically coupled in radiative transfer. To disentangle their individual effects, we fix the radiance ratio (\(I_{\rm ratio}\)) by simultaneously adjusting $\rm \frac{1}{\lambda} \frac{1}{T^{2}}$ and \(\frac{dT}{d\log P} \), allowing a controlled, quantitative analysis of their impact on polarization.

\begin{figure}
	\includegraphics[width=1\columnwidth]{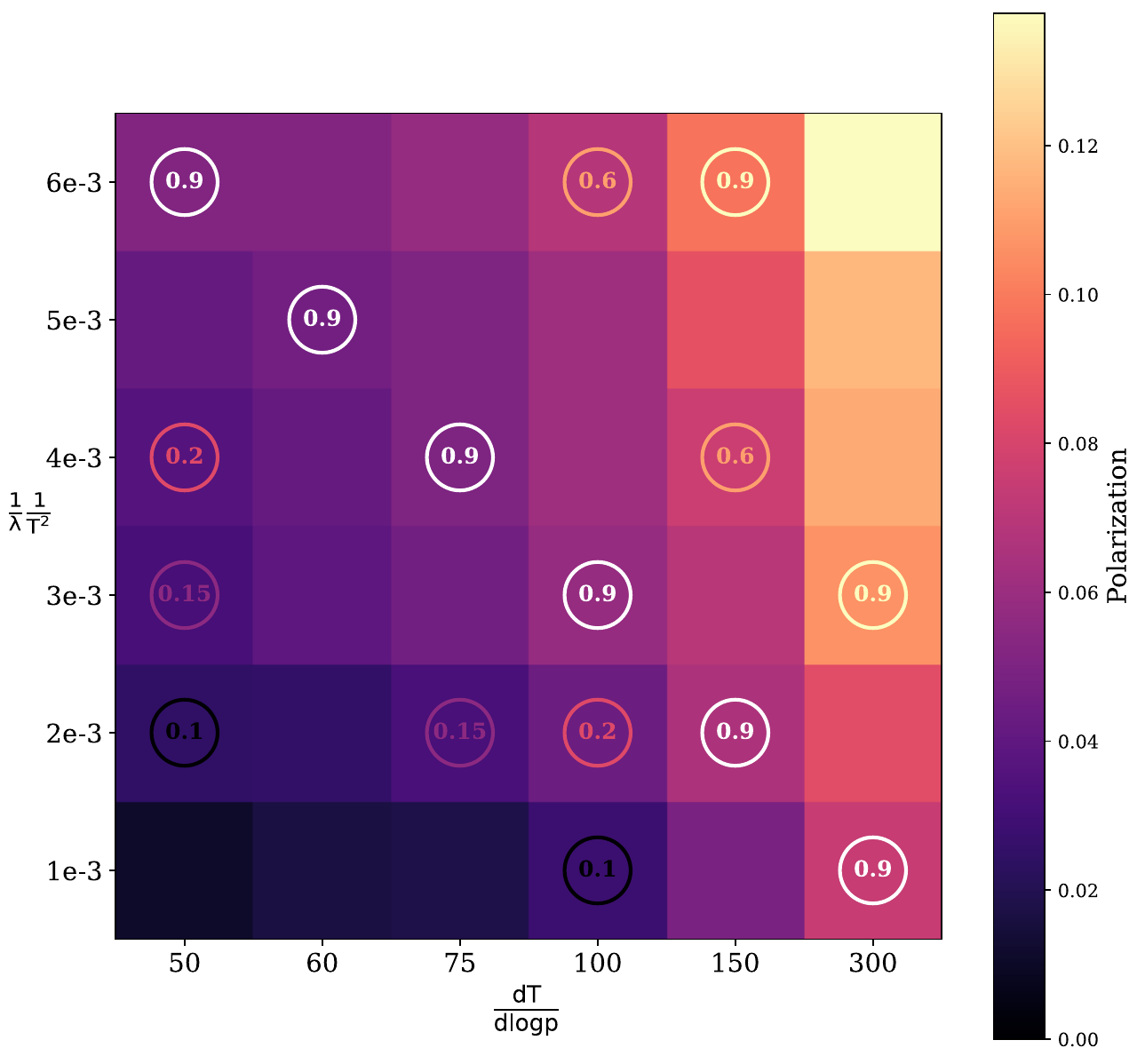}
    \caption{Linear polarization for atmospheres with a high-altitude cloud is shown as a function of wavelength-dependent scaling ($\frac{1}{\lambda} \frac{1}{T^{2}}$) and temperature gradient ($ \frac{dT}{d\log P}$). Note the x axis is not a linear scale, only the grid points list in the x and y axis are considered in this plot. The color-coded circles represent the the product of $\frac{1}{\lambda} \frac{1}{T^{2}}$ and $\frac{dT}{d\log P}$. When circle value remains is the same, the polarization is similar. These two factors jointly determine the $ I_{\rm ratio}$, which governs the single scattering polarization. A larger product of $\frac{1}{\lambda} \frac{1}{T^{2}}$ and $\frac{dT}{d\log P}$ leads to stronger single scattering polarization.}
    \label{fig5}
\end{figure}

As illustrated in Figure \ref{fig5}, the polarization degree is highly sensitive to both wavelength and temperature gradient. Specifically, \(P\) increases approximately linearly with both $\frac{1}{\lambda} \frac{1}{T^{2}}$ and \(\frac{dT}{d\log P} \). Physically, this reflects the increasing anisotropy in the radiation field: larger values of either parameter enhance the vertical flux relative to the oblique component, favoring photon scattering at near-perpendicular angles. At the extreme case of a zero temperature gradient, the radiation field becomes nearly isotropic around the scatterers, even at the limb. Consequently, the net polarization of radiation observed would approach zero. Thus, polarization decreases with increasing wavelength (due to the shift of thermal emission to longer wavelengths) but increases with steeper temperature gradients.

The color-coded circles at the grid points in Figure \ref{fig5} quantitatively map the relationship between polarization degree \(P\) and the two governing parameters. Notably, when the product \( \frac{1}{\lambda} \frac{1}{T^{2}} \cdot \frac{dT}{d\log P} \) remains constant, the resulting polarization degree is nearly unchanged. This highlights that thermal polarization is particularly sensitive to the local temperature gradient at the atmospheric layer where the optical depth reaches unity. Building on this relationship, we will explore in the next section how thermal radiation polarization can be used as a diagnostic tool to constrain the temperature structure of the atmosphere.

\subsection{The Effect of Molecular Absorption}\label{sec:4.4}

\begin{figure}
	\includegraphics[width=1.02\columnwidth]{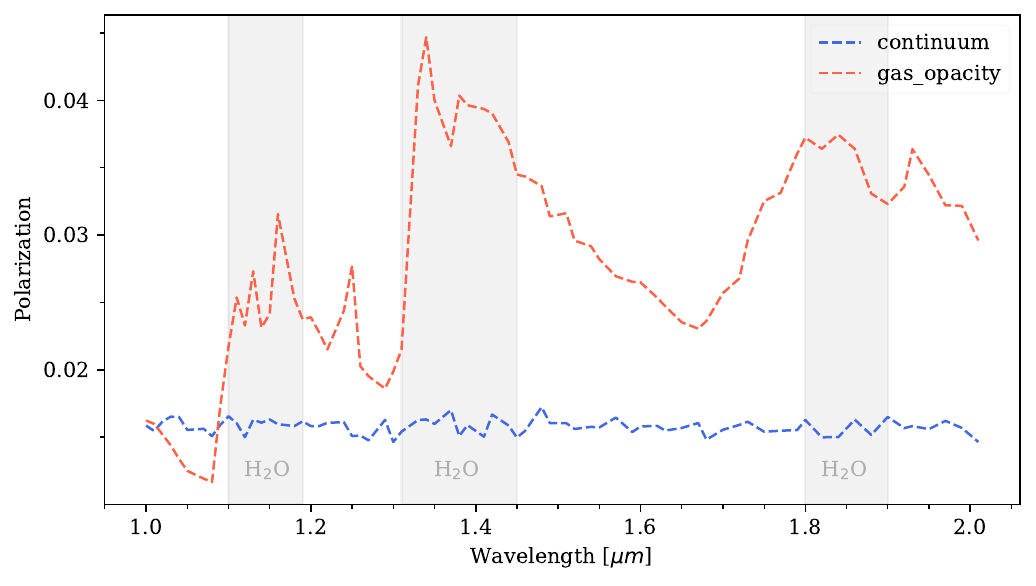}
    \caption{The effect of atmospheric molecular absorption on the polarization spectrum at the limb. Blue curves shows the the wavelength response of the polarization spectrum only considering continuum absorption. When Considering the selective absorption of water vapor molecules, the polarization is significantly increased in the band of water vapor molecule absorption, which is because the molecular absorption affects the atmospheric height of $\tau=1$ layer, and the temperature corresponding to this atmospheric height and the temperature gradient directly affect the polarization.}
    \label{fig6}
\end{figure}

Figure \ref{fig6} presents simulated polarization spectra in the near-infrared for two cases: (1) with only continuum absorption and (2) with both continuum and molecular absorption from water vapor. The blue curve shows the polarization spectrum in the continuum-only scenario, where the degree of polarization remains low. This is because low continuum opacity allows thermal radiation to emerge from deeper, hotter atmospheric layers. At these depths, the high temperatures suppresses the polarization even with the presence of a relative large temperature gradient.

By contrast, when water vapor absorption is included (red curve), a notable increase in polarization is observed at water absorption bands centered around 1.1, 1.4, and 1.9 \(\mu m\) (indicated by the shaded gray regions). These enhancements arise because the increased opacity in absorption bands shifts the effective photosphere to higher altitudes, where temperatures are lower. Given a positive temperature gradient, this vertical displacement results in a stronger anisotropy in the local radiation field. Specifically, the radiance ratio \(I_{\rm ratio}\) at \(\tau = 1\) is greater in absorption bands than in adjacent continuum regions, i.e., \(I_{\rm ratio}^{\rm gas} \geq I_{\rm ratio}^{\rm con}\). This increased anisotropy leads to higher polarization in absorption bands.

It is important to note that these results are based on an idealized atmosphere with a constant positive temperature gradient. In reality, substellar atmospheres often exhibit more complex thermal structures, including variable gradients and possible temperature inversions (\( dT/d\log P < 0\)) \citep{burrows1999chemical, zhang2020atmospheric, faherty2024methane}. Therefore, whether molecular absorption enhances or suppresses polarization depends on both the temperature and its gradient at the atmospheric layer where \(\tau = 1\). In general, polarization increases when the product of \(\frac{1}{\lambda} \frac{1}{T^{2}} \cdot \frac{dT}{d\log P}\) is greater in the absorption band than in the adjacent continuum; otherwise, it decreases.
The impact of molecular absorption on thermal polarization is indirect as it arises from shifts in the probed atmospheric layer, where local temperature and gradient dictate the resulting polarization degree.

\begin{figure*}
\includegraphics[width=2.15\columnwidth]{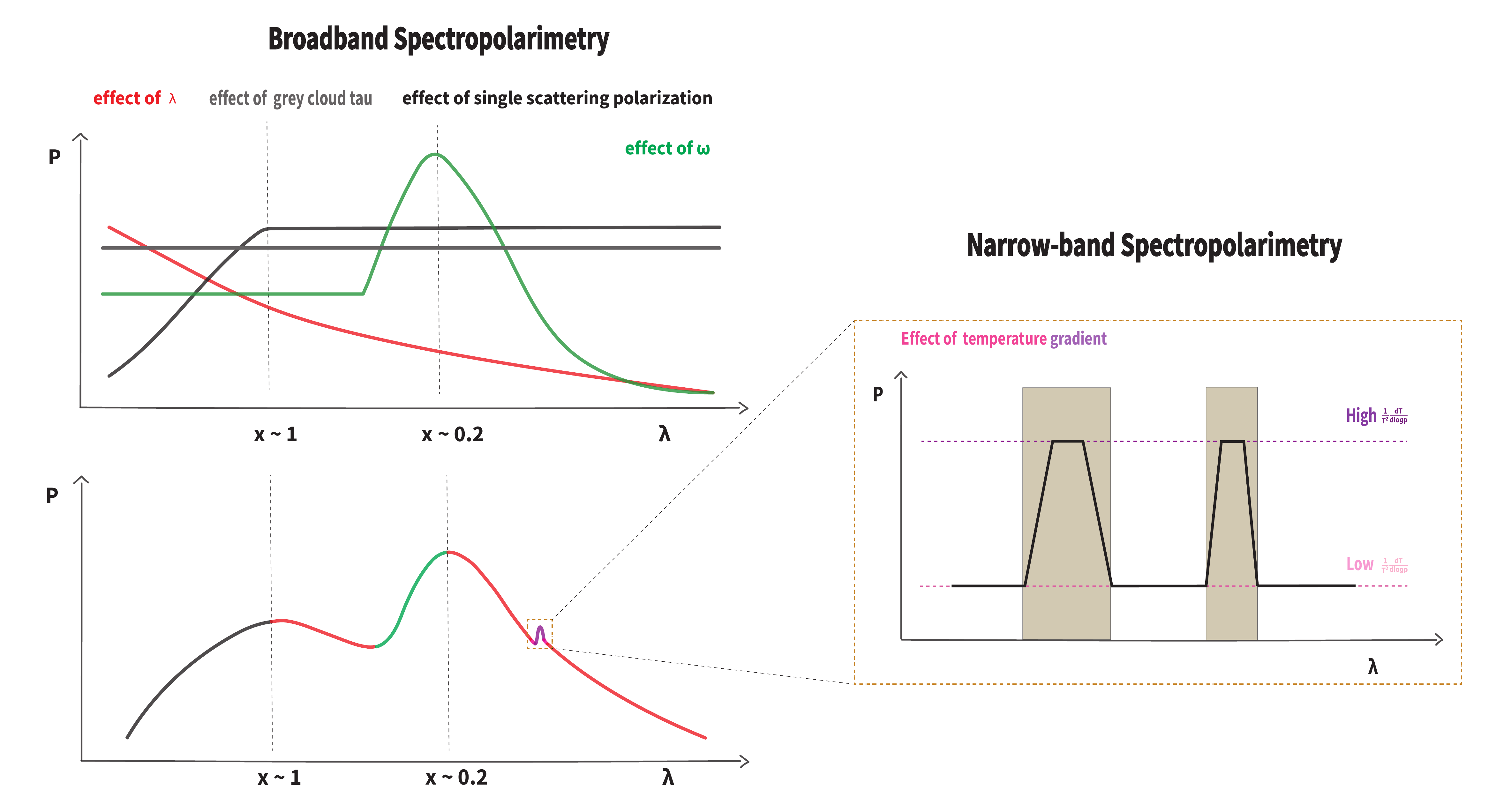}
    \caption{Schematic figure illustrates how various factors collectively influence the thermal polarization spectrum. The left panel summarizes the effect of single scattering polarization, cloud single-scattering albedo, gray cloud optical thickness and wavelength that combine to shape the broadband polarization spectrum, potentially creating a distinctive two-peak feature in the bottom left panel. The right panel demonstrates the temperature structure shape the fine polarization in the molecular absorption band. Constant temperature gradient with pressure are considered.
    Purple and pink lines represent 
    high and low \(\rm \frac{1}{T} \frac{dT}{d \log p} \) respectively.
    }
    \label{fig7}
\end{figure*}

\section{Implications of Spectropolarimetry for Atmospheric Characterization}\label{sec:5}

In the preceding sections, we analyzed how key atmospheric parameters of cloud particle size, cloud optical thickness and temperature structure individually affect thermal radiation polarization. Here, we explore how these parameters interact and collectively shape the polarization spectrum. By understanding their combined influence, we aim to demonstrate how future observations of thermal polarization spectra can be used to constrain the atmospheric properties of brown dwarfs.

\subsection{Combined Effect of Determining Factors on Shaping the Thermal Polarization Spectrum}\label{sec:5.1}

As shown in Figure \ref{fig4}, single scattering polarization, cloud single-scattering albedo, and cloud optical thickness collectively govern the broadband behavior of the polarization spectrum. Building on this, we present a schematic in Figure \ref{fig7} to illustrate how these factors interact to shape the thermal polarization spectrum. The red curve in Figure \ref{fig7} represents the polarization response with the wavelength, $\lambda$, as discussed in Section~\ref{sec:4.3}. For clarity, we assume a constant $\tau_{\rm cld}$ to isolate the key trends in the convolved polarization spectrum, as the dependence of cloud optical thickness is stronger, and tend to
dominate the broadband trend. 

Together, the effects of single scattering polarization, albedo, and wavelength shift create a characteristic broadband two-peak structure in the polarization spectrum. The first peak arises when the size parameter $x \sim 1$, where single scattering polarization is maximized. The second peak is shaped by the single-scattering albedo, typically around $x \sim 0.2$, where absorption suppresses multiple scattering and enhances net polarization. This two-peak structure is a robust signature across a range of cloud species, assuming a transition from Mie to Rayleigh scattering with wavelength and a wavelength-dependent albedo. Although the exact value of $x$ at which $\omega$ peaks may differ between cloud types, the overall broadband shape remains consistent. Importantly, the positions of these peaks shift with wavelength as the effective particle radius varies, offering a promising diagnostic of cloud particle size in brown dwarf atmospheres.

In contrast, narrow-band features in the polarization spectrum are more sensitive to wavelength-dependent molecular absorption and the vertical temperature structure near the photosphere. These factors introduce fine-scale structure in the polarization spectrum, as illustrated in the bottom panel of Figure \ref{fig7}. The brown dashed box highlights a polarization peak caused by molecular absorption. The strength of this feature depends on the contrast in the product of temperature and vertical temperature gradient, expressed as $\frac{1}{T^{2}} \frac{dT}{d\log P}$, between the absorption band and the adjacent continuum. As shown in the zoom-in plot on the right, the purple and pink curves represent vertically constant temperature gradients with different $\frac{1}{T^{2}} \frac{dT}{d\log P}$. The molecular absorption band probes higher atmospheric layers (shaded region), where the  $\frac{1}{T^{2}} \frac{dT}{d\log P}$ is larger. A stronger $\frac{1}{T^{2}} \frac{dT}{d\log P}$  in the absorption band leads to enhanced polarization, whereas a smaller one in the continuum suppresses it. Consequently, molecular bands typically exhibit enhanced polarization relative to the surrounding continuum, since they probe higher atmospheric layers where, under a constant positive temperature gradient, the temperature is lower. This contrast creates distinct polarization peaks within the absorption features.

\subsection{Constraints on Cloud Grain Size via Broadband Spectropolarimetry}\label{sec:5.2}

In this section, we investigate the potential of thermal emission spectropolarimetry to constrain cloud grain sizes. To validate the predicted broadband two-peak polarization feature (as schematically illustrated earlier), we simulate the polarization spectrum using the ARTES radiative transfer code. We assume a gray cloud and constant molecular absorption across the entire wavelength range (0.6–6 \(\mu\)m). The cloud optical thickness is fixed at unity, and molecular absorption is uniformly set to its value at 0.6 \(\mu\)m to eliminate narrow-band variations from the cloud optical depth, temperature gradient, and absorption features.

\begin{figure}
	\includegraphics[width=1.02\columnwidth]{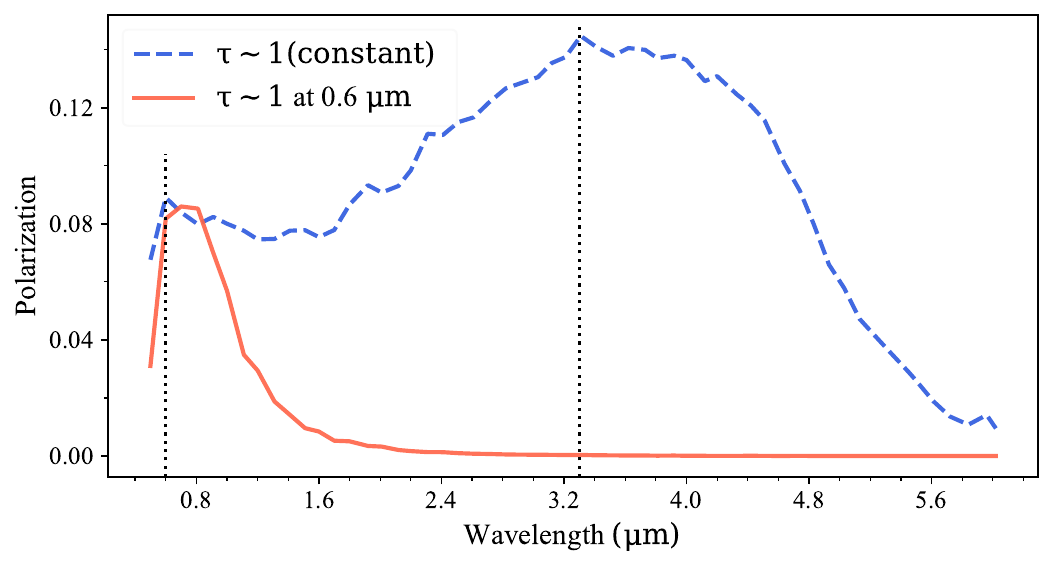}
   \caption{The blue dashed curve represents the broadband two-peak polarization feature, modeled under the assumption of a wavelength-independent cloud optical depth. The simulation produces two polarization peaks at $x \sim 1$ and $x \sim 0.2$, consistent with the theoretical schematic. The red curve accounts for wavelength-dependent cloud optical depth, where the cloud optical thickness diminishes in the near-infrared. As a result, only the first peak (at $x \sim 1$) in the optical band remains observable.}
    \label{fig8}
    \end{figure}

Figure \ref{fig8} presents the simulated polarization spectrum at the limb of a brown dwarf with a high-altitude \(\rm MgSiO_3\) cloud layer. As expected, the blue curve exhibits two polarization peaks: one near 0.6 \(\mu\)m and the other around 3.2 \(\mu\)m. These correspond to \(x \sim 1\) and \(x \sim 0.2\), respectively, for \(\rm MgSiO_3\) clouds with an effective particle radius of 0.1 \(\mu\)m. The red curve shows the case with wavelength-dependent cloud optical thickness. According to the right panel of Figure \ref{fig3}, the extinction cross-section of \(\rm MgSiO_3\) drops significantly with wavelength in this range. Consequently, when the cloud optical depth is normalized to unity at 0.6 \(\mu\)m, it becomes negligible at longer wavelengths which will effectively eliminating the second polarization peak due to reduced photon scattering.

The visibility of both polarization peaks depends on maintaining an optimal cloud optical depth across the relevant wavelengths. Brown dwarf and exoplanet atmospheres span a wide range of temperatures, leading to condensation of different cloud species. These species, along with their particle size distributions, influence the wavelength dependence of the cloud extinction cross-section. If the an optimal depth can sustain across a wide spectral range (i.e., gray clouds with $\tau \sim 1$ ), both two polarization peaks can be observed. Otherwise, wavelength-dependent cloud optical thickness may cause one or both peaks to vanish due to insufficient or excessive scattering at certain wavelengths.

\begin{figure}
	\includegraphics[width=1.02\columnwidth]{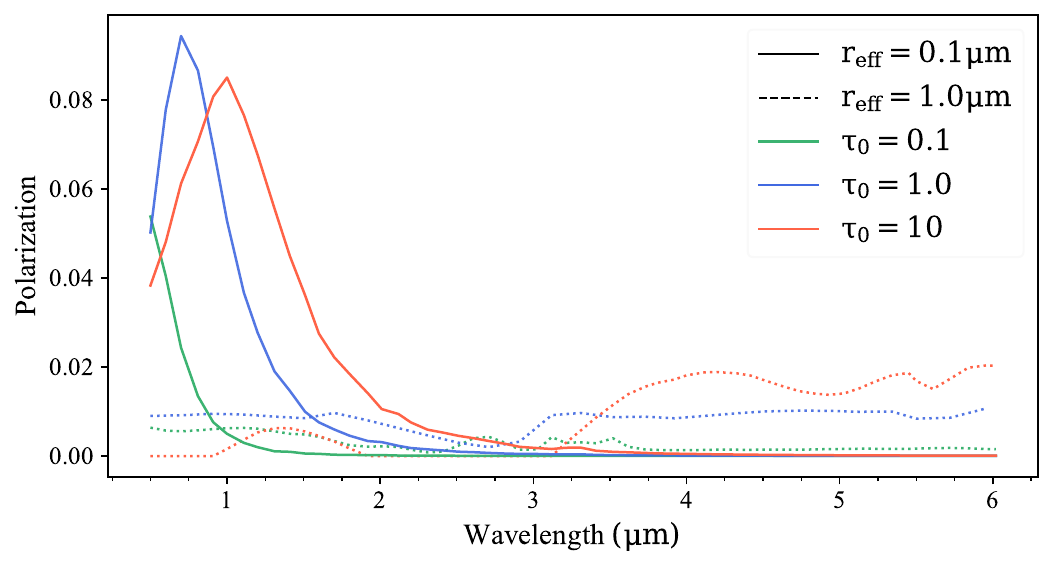}
     \caption{ The effect of wavelength-dependent cloud optical depth on the polarization peak and its position is analyzed by considering wavelength-dependent cloud optical thickness. The optical depth at 0.5 $\mu$m ($\tau_0$) serves as a reference. Two cloud effective radii ($r_{\rm eff}$) of 0.1 $\mu$m and 1 $\mu$m are examined. For $r_{\rm eff} = 0.1 \, \mu$m, the polarization peak shifts to longer wavelengths as the cloud optical thickness increases. In contrast, for larger cloud particles ($r_{\rm eff} = 1 \, \mu$m), no significant linear polarization is observed in the near-infrared.
    }
    \label{fig9}
\end{figure}

Another important diagnostic feature is the wavelength shift of the polarization peak with varying cloud particle sizes. Figure \ref{fig9} illustrates this dependence by comparing polarization spectra for \(\rm MgSiO_3\) clouds with effective radius of 0.1 and 1 \(\mu\)m. To account for the strong wavelength dependence of extinction, we normalize the cloud optical depth to 0.1, 1, and 10 at 0.6 \(\mu\)m, allowing it to vary with wavelength according to the extinction slope shown in Figure \ref{fig3}. For \(r_{\rm eff} = 0.1\ \mu\)m, the first polarization peak appears near 0.628 \(\mu\)m and remains visible across all optical thickness cases. As the cloud optical depth increases (up to \(\tau = 10\)), this peak shifts toward longer wavelengths due to the enhanced contribution of multiple scattering. Cases with higher optical depths are excluded, as Monte Carlo simulations tend to underestimate flux in such regimes \citep{krieger2024improving}. In contrast, for \(r_{\rm eff} = 1\ \mu\)m, no distinct polarization peak is observed from the optical to infrared. While the first peak theoretically lies in the mid-infrared near 6.28 \(\mu\)m, the rapidly decreasing extinction cross-section at longer wavelengths reduces scattering efficiency, weakening the polarization signal. Nevertheless, polarization still increases toward longer wavelength, especially in the higher optical depth cases (dotted curves), which reflects the underlying size parameter dependence.

The sensitivity of polarization peaks to the effective radius of cloud particles highlights the potential of broadband spectropolarimetry as a diagnostic for cloud grain size in brown dwarf atmospheres. For instance, small silicate grains often exhibit vibrational absorption features near \(\sim10\ \mu\)m. Importantly, the grain size constraints obtained through polarization are independent of traditional spectroscopic methods and may be more robust than those inferred from Mie scattering features in the flux spectrum alone.

\subsection{Constraints on the Temperature Structure via Narrow-Band Spectropolarimetry}\label{sec:5.3}

\begin{figure}
	\includegraphics[width=1.02\columnwidth]{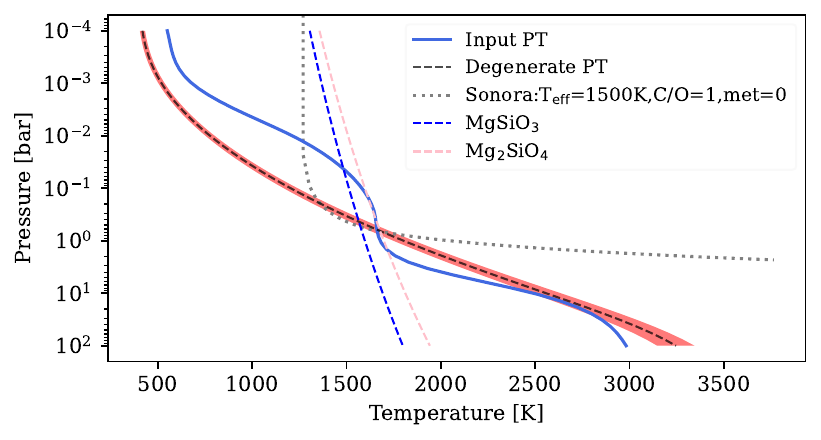}
    \caption{A comparison of the median retrieved thermal profiles (black curve) with the input $P\text{--}T$ profile (blue curve). The red-shaded region represents the 1$\sigma$ and 2$\sigma$ central credible intervals around the median profile. The input and degenerate $P\text{--}T$ profiles exhibit notably different temperature gradients but, when paired with varying gases abundances, produce similar flux spectra.
    A Sonora grid $P\text{--}T$ profile with $T=1500K$, C/O=1, met=1 is shown with gray dotted line. The adiabatic slope in the deep atmosphere is much steeper than our toy model.}
    \label{fig10}
\end{figure}

The sensitivity to the temperature gradient that we studied here can provide unique evidence in constraining the T-P profile. It may particularly be useful in resolving the degeneracy between gas abundances and the pressure–temperature structure only from the intensity spectra as demonstrated in previous study \citep{morley2014water,burrows2006and,ackerman2001precipitating,burningham2017retrieval,molliere2020retrieving,tremblin2015fingering,barrado202315nh3}. In this section, we investigate how narrow-band spectropolarimetry can be used to probe and constrain atmospheric temperature structures.

\begin{figure*}
	\includegraphics[width=2.15\columnwidth]{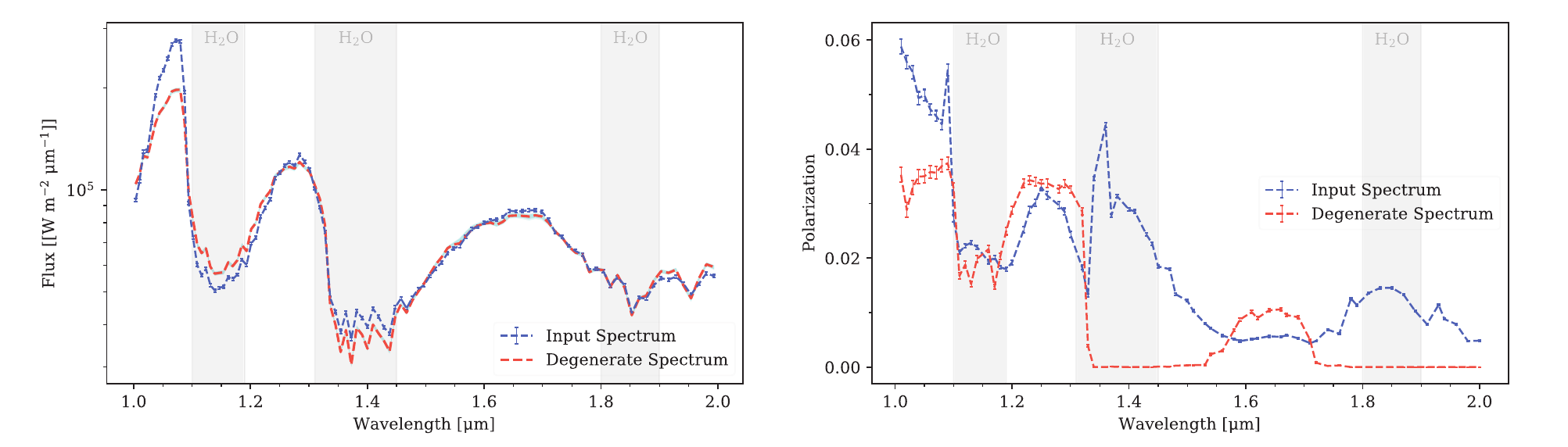}
    \caption{Left panel: Top-of-atmosphere flux spectra generated from two degenerate $P\text{--}T$  profiles with differing chemical abundances. The input flux spectrum, simulated with S/N = 100, is shown as the blue dashed curve with error bars. The red dashed curve represents the retrieved median spectrum, and the cyan shaded regions denote the 1$\sigma$ and 2$\sigma$ confidence intervals. Right panel: Corresponding polarization spectra for the two cases. Uncertainties estimated from Monte Carlo radiative transfer simulations using 10$^{9}$ photon packages, are indicated by error bars. The polarization spectra exhibit pronounced differences, particularly within molecular absorption bands.}
    \label{fig11}
\end{figure*}

We start with a toy model of a cool, cloudy brown dwarf to demonstrate how spectropolarimetry can help break degeneracies between gas abundances and $P\text{--}T$  structure. The temperature profile (blue curve in Figure \ref{fig10}) is parameterized using the 13-knot method of \citet{line2015uniform}. This approach employs 13 knots to define the basic temperature structure, and then interpolate the temperatures across all considered pressure layers. In this setup, we introduce a nearly isothermal region between 0.1 and 1 bar. This abrupt change in the temperature gradient, occurring within the dominant flux-contributing pressure range, will significantly impact the polarization spectrum. The model includes opacities from four dominant gases: H\textsubscript{2}O, CO, CO\textsubscript{2}, and CH\textsubscript{4}, all of which influence the near-infrared polarization features. To provide a source of scattering, we include a single high-altitude MgSiO\textsubscript{3} cloud layer at \(P = 10^{-3} \) bar, with a gamma particle size distribution (\( r_{\rm eff} = 0.1\ \mu\mathrm{m} \), \( v_{\rm eff} = 0.05 \)) and a constant optical depth of \( \tau = 1 \) across all wavelengths.

The left panel of Figure \ref{fig11} shows the corresponding emission spectrum, calculated using PICASO \citep{batalha2019exoplanet, mukherjee2023picaso}. Using this as mock data, we perform atmospheric retrieval with a different $P\text{--}T$  parameterization \citep{madhusudhan2018atmospheric}, which allows for thermal inversions. The retrieved model parameters and their corresponding prior ranges are listed in Table~\ref{tab:priors}. The retrieved temperature profile (black curve in Figure \ref{fig10}) deviates significantly from the input profile, particularly between 0.1 and 1 bar, where it exhibits a steep gradient instead of the expected isothermal structure. The upper atmosphere also shows lower temperatures in the retrieved case, though the gradient is comparable. Cloud properties also deviate substantially from the input values. The retrieved cloud layer is deeper, with a base pressure of \( \log P = -0.34^{+0.03}_{-0.02} \) dex, a vertical extent of \(\log dp = 0.10^{+0.05}_{-0.04} \) dex, and a higher optical depth of \( \tau = 2.88^{+0.01}_{-0.01} \). Gas abundances shift as well: H\textsubscript{2}O and CH\textsubscript{4} are overestimated in the input, while CO is underestimated by ~5 orders of magnitude in the degenerate solution (see Appendix A.2).

\begin{figure*}
	\includegraphics[width=2.2\columnwidth]{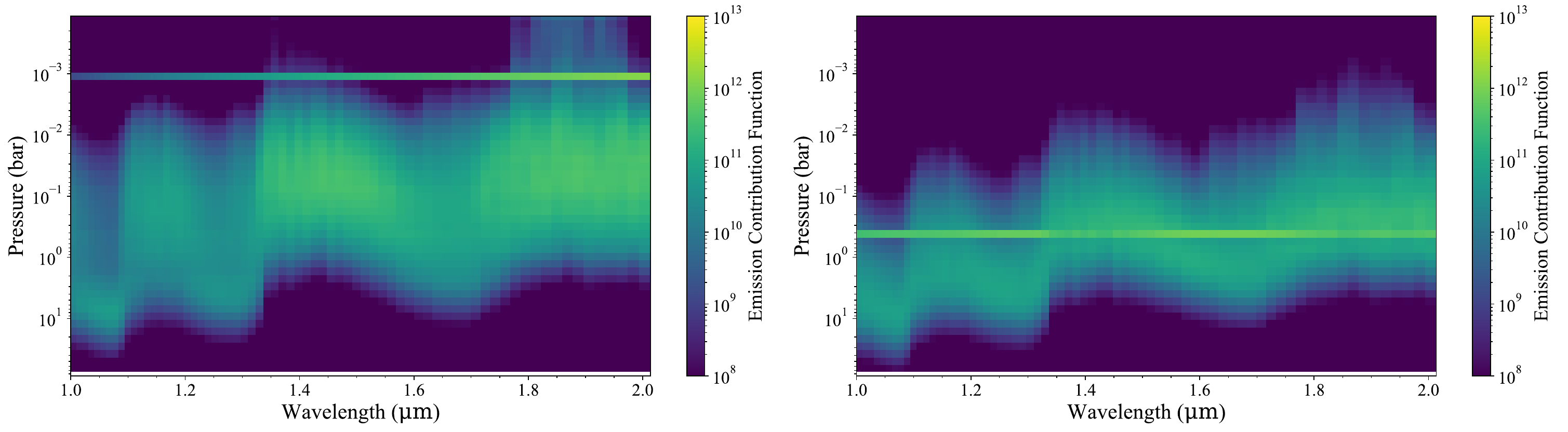}
    \caption{Flux contribution as a function of wavelength and pressure is shown for two atmospheric scenarios: the input $P\text{--}T$ profile (left panel) and the degenerate $P\text{--}T$ profile (right panel). For the degenerate scenario, the flux predominantly originates from deeper atmospheric layers (due to stronger water absorption, see Figure ~\ref{fig:abun}) with a steeper temperature gradient.}
    \label{fig12}
\end{figure*}

The right panel of Figure \ref{fig11} shows the polarization spectra for both the input and degenerate atmospheric profiles using ARTES. It shows that although the emission spectra look similar, the polarization spectra are remarkably different at certain wavelength ranges. In the $\rm H_{2}O$ band around 1.1 $\rm \mu m$, both scenarios show comparable polarization signals. However, at wavelengths near 1.4 and 1.9 $\rm \mu m$, the polarization trends between the input and degenerate cases are completely opposite. The polarization signal in the $\rm H_{2}O$ band center around 1.1 $\rm \mu m$ shows similar polarization signal, while around 1.4 and 1.9 $\rm \mu m$ shows completely opposite trend between input and degenerate case. To further illustrate the origins of these differences, Figure \ref{fig12}  shows the flux contribution as a function of wavelength and pressure. These plots highlight the pressure levels that contribute most significantly to the flux at each wavelength. As shown in Figure \ref{fig11}, for the input $P\text{--}T$  profile scenario, the molecular absorptions shift the photosphere to higher altitudes compared to adjacent continuum bands. For instance, in the $\rm H_{2}O$ band at 1.1 \(\mu\)m, flux originates from 0.1–1 bar, within the isothermal region which suppresses polarization. Conversely, bands like the 1.4 and 1.9 \(\mu\)m $\rm H_{2}O$ features probe higher altitudes with steeper temperature gradients, producing polarization peaks.

In the degenerate case, the polarization trends are reversed. The photosphere in molecular bands rises above the cloud top, diminishing polarization. Meanwhile, the large temperature gradient in deeper layers enhances polarization in adjacent continuum bands. Despite producing nearly identical emission spectra with sightly difference at short wavelengths, however, the two models yield significant different polarization spectra. These discrepancies reflect differences in temperature structure and cloud vertical location, which may not be discernible from flux data alone.

Figure \ref{fig11} highlights how spectropolarimetry can reduce degeneracies between temperature and abundance retrievals in simplified models. In actual brown dwarf and exoplanetary atmospheres, changes in the temperature gradient may not be as extreme but could still involve temperature perturbations. For instance, hot Jupiters may exhibit thermal inversions in their upper atmospheres or isothermal deep layers due to stellar irradiation \citep{sudarsky2003theoretical, fortney2008unified}. While isolated brown dwarfs are not expected to show such features, they could still develop temperature inversions  due to additional astrophysical processes such as auroral activity  \citep{hallinan2015magnetospherically, faherty2024methane}, introducing localized changes in vertical temperature gradients. These perturbations will imprint themselves in polarization spectra, particularly within molecular bands, offering a powerful and complementary tool to flux-based diagnostics for probing atmospheric thermal structure.

\section{Discussion and Conclusions}\label{sec:6}

In previous sections, we examined how fundamental atmospheric properties influence thermal spectropolarimetry and identified key schematic features that can constrain cloud particle sizes and break degeneracies in flux-based retrievals between temperature structure and gas abundances. Here, we build upon those results to discuss additional insights from our theoretical exploration of these contributing factors and evaluate the detectability of the predicted polarization features with current spectropolarimetry.

\subsection{Polarization Dependence on Effective Temperature}\label{sec:6.1} 

A noteworthy byproduct of this study is the dependence of thermal polarization on the effective temperature of substellar objects. \citet{sanghavi2018photopolarimetric} found that cooler brown dwarfs tend to exhibit stronger polarization than hotter ones when assuming identical atmospheric conditions and oblateness. This effect is closely tied to gravitational darkening, a phenomenon arising from rapid rotation, which causes brown dwarfs to become oblate and develop latitudinal variations in surface gravity and temperature. According to classical gravitational darkening theory \citep{von1924radiative, lucy1967gravity, claret2000studies, reiners2003effects, lara2011gravity}, reduced effective gravity at the equator leads to lower temperatures and dimmer emission relative to the hotter, brighter poles. This latitudinal temperature gradient enhances atmospheric asymmetry, thereby increasing the net polarization signal.

\begin{figure}
	\includegraphics[width=1\columnwidth]{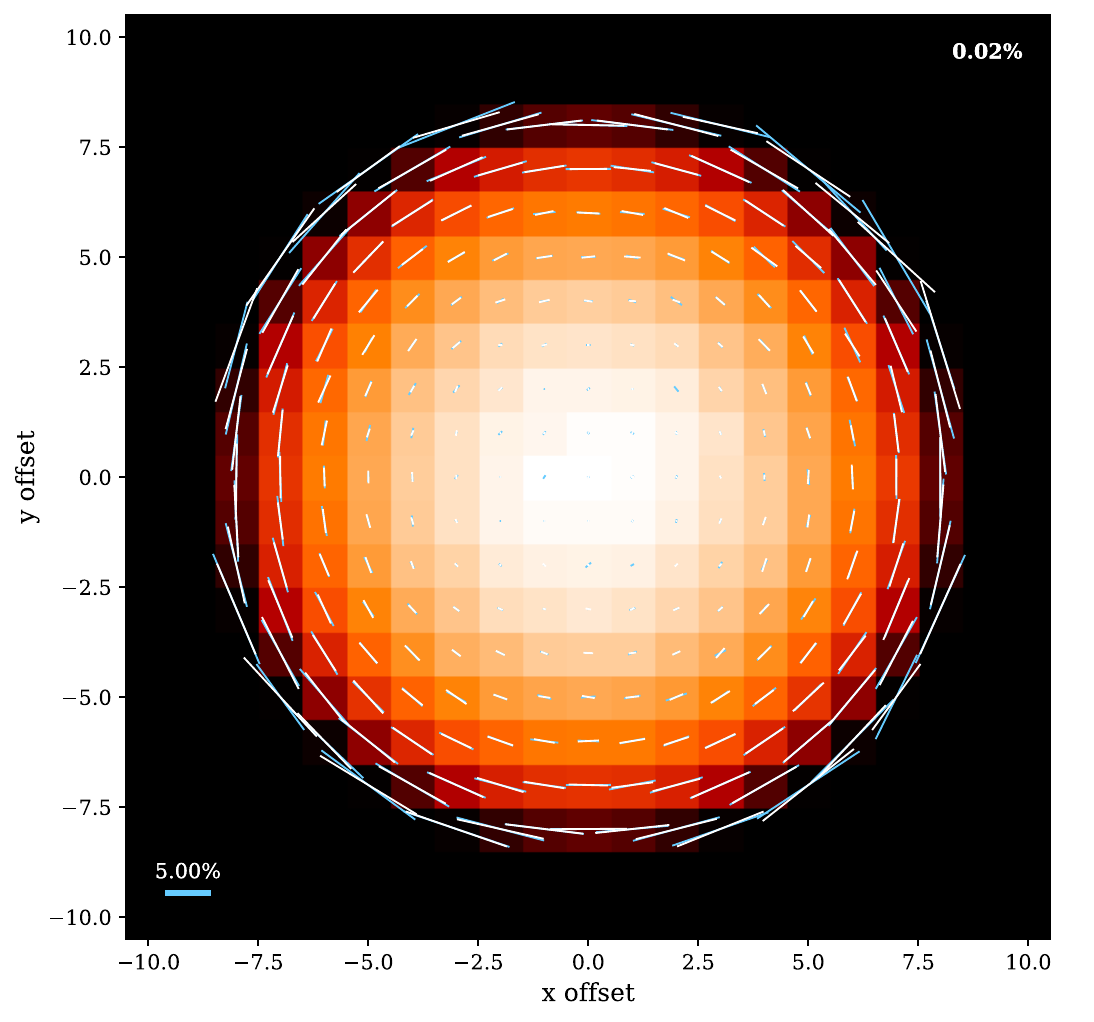}
    \caption{Influence of effective temperature on thermal polarization. Blue and white vectors represent polarization at the limb for atmospheres with \(T_{0.1\,\text{bar}} = 1500\,\text{K}\) and \(1800\,\text{K}\), respectively. Limb polarization is stronger in the cooler atmosphere, illustrating the temperature dependence of polarized thermal emission.}
    \label{fig13}
\end{figure}

From a disk-resolved perspective, our simulations also show that limb polarization becomes more pronounced as the effective temperature decreases. Figure \ref{fig13} presents polarization maps at 1.5 $\mu$m for two brown dwarf models with effective temperatures of 1800 K and 1500 K, respectively. All other parameters, including cloud properties and temperature gradients, are held constant. The cooler object exhibits a stronger polarization signal at the limb ($\sim$9\%) compared to the hotter case ($\sim$7\%). This trend aligns with the temperature dependence described by the polarization scaling relation \(\sim \frac{1}{\lambda}\frac{1}{T^{2}} \cdot \frac{dT}{d\log P}\). For a fixed vertical temperature gradient, a lower temperature increases the contrast between vertically and obliquely emitted radiation, enhancing the observed polarization.

These results suggest that cooler, fast-rotating brown dwarfs are especially promising targets for thermal spectropolarimetry. Rapid rotation leads to equatorial bulging and significant latitudinal temperature differences \citep{sengupta2010observed}, while low effective temperatures naturally increase polarization sensitivity. When combined with cloud-induced variability observed in time-series studies \citep{wang2021diurnal}, can lead to amplified overall polarization signal, potentially making it more detectable. Another promising target group is the spatially resolved direct-imaging planet. As noted by \citet{de2011characterizing}, planets like those in the HR 8799 system are more likely than hotter brown dwarfs to show detectable levels of polarization due to their cooler, cloudier, and more oblate atmospheres \citep{marley2011probing, madhusudhan2011model}. These planets often lie in temperature regimes where cloud formation and thermal emission are most conducive to generating polarized light, making them excellent candidates for future spectropolarimetric observations.

\subsection{Detectability of Predicted Features in Net Polarization Spectrum}\label{sec:6.2}

This study primarily focuses on disk-resolved polarization, particularly at the planetary limb where cloud scattering produces the strongest near-infrared signal. However, directly resolving the surfaces of brown dwarfs is not feasible with current or near-future imaging technologies. For unresolved point sources, the disk-integrated polarization from thermal emission is typically negligible unless the atmosphere departs from spherical symmetry. Such asymmetries can arise from rotation-induced oblateness, heterogeneous cloud bands or spots, day–night temperature contrasts, or even the presence of rings \citep{sengupta2010observed, marley2011probing}. \citet{de2011characterizing} demonstrated that, under favorable conditions, such deviations can yield net polarization levels exceeding 0.1\%, and occasionally reaching several percent in the near-infrared.

\citet{stolker2017polarized} investigated the impact of latitudinal and longitudinal cloud variations, circumplanetary disks, atmospheric flattening, and particle properties on the integrated polarization signal. They found that while limb polarization can be significant, spatial averaging typically reduces the net disk-integrated signal by about an order of magnitude; for example, a case with banded clouds showed a limb polarization near 15\%, but the net value dropped to just 1.33\%. This scenario assumes relatively extreme cloud distributions, with thick clouds between \(-30^\circ\) and \(30^\circ\) latitudes and thinner clouds at higher latitudes, yielding a limb-to-net scaling factor of $\sim$ 0.1. Oblateness, induced by rapid rotation, can enhance net polarization. Rotational flattening introduces large-scale atmospheric asymmetry, thereby boosting the disk-integrated signal. The dependence of polarization on oblateness, especially in single scattering regimes (\(x \leq 1.5\)), has been modeled analytically by \citet{dolginov1995propagation}, \citet{simmons1982analytic}, and \citet{sengupta2001probing}, all showing a monotonic increase in net polarization with increasing oblateness.

\begin{figure*}
	\includegraphics[width=1.4\columnwidth]{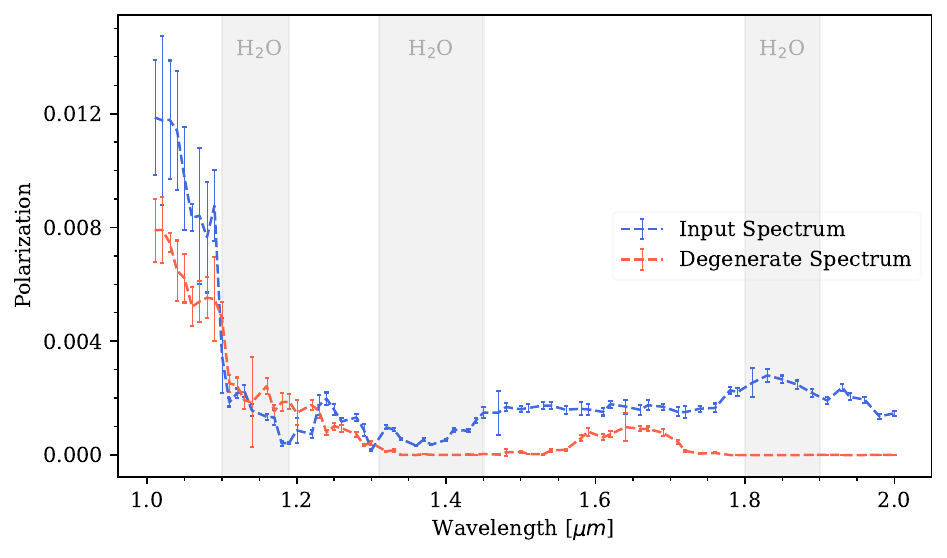}
    \caption{Net polarization spectra for two degenerate scenarios that differ in their $P\text{--}T$  profiles and chemical abundances. The polarization feature differences highlighted in Figure~\ref{fig11} remain discernible in the net polarization spectra.}
    \label{fig14}
\end{figure*}

In Section~\ref{sec:5.3}, we showed that disk-resolved polarization within the water absorption bands can serve as a diagnostic of thermal structure and cloud vertical distribution. Extending this analysis to the disk-integrated case, we adopt the same atmospheric configuration but introduce a large-scale asymmetry, with thick cloud ($\tau=5$) distribute clouds between \(-30^\circ\) and \(30^\circ\) latitudes ($\tau=1$) and thinner clouds at higher latitudes. Figure~\ref{fig14} shows the resulting disk-integrated polarization spectra for both the nominal and degenerate atmospheric cases. Compared to the disk-resolved polarization, the overall net signal is reduced by more than an order of magnitude, reaching $\sim10^{-3}$ within the water absorption band for the nominal model. Although muted by disk averaging, the differences between the two scenarios remain distinguishable, as highlighted by the shaded regions in the figure. This configuration represents an optimistic case with an atmosphere with strong disk-resolved polarization and substantial global asymmetry, yielding a net near-infrared polarization of ($10^{-3} -10^{-2}$). In more realistic cases, the expected level is around $10^{-4}$. For reference, the detected linear polarization of brown dwarfs Luhman 16A and B is \(P_{\rm A} = 0.031\%\) and \(P_{\rm B} = 0.01\%\), respectively \citep{millar2020detection}, consistent with atmospheric models featuring cloud bands. In more symmetric or temporally variable systems, low disk-resolved polarization combined with randomized global cloud and temperature structure distribution may lead net polarization may fall below $10^{-5}$.

Assuming a realistic net polarization of $10^{-4}$ in the near-infrared, we assess the feasibility of detecting such signals with current spectropolarimetry. For instance, Luhman 16A has been observed in the H band with the VLT/NaCo imager \citep{lenzen2003naos, rousset2003naos}, which lacks spectropolarimetric capability. The CFHT instrument SPIRou \citep{donati2020spirou} provides high-resolution (R $\sim$ 70000) spectropolarimetry over 0.98–2.35 $\mu$m with a polarimetric sensitivity of 10 ppm. However, SPIRou’s high-resolution linear spectropolarimetry is strongly signal-to-noise limited for faint near-infrared sources such as late-L and T-type brown dwarfs ($J \approx$ 13–17 mag). Detecting $10^{-4}$-level polarization requires flux S/N $\ge$ 10000, which is infeasible even for the nearby Luhman 16 system.

A more practical approach is low-resolution spectropolarimetry (R $\sim$ 100), achieved by combining thousands of high-resolution pixels to boost S/N. Using SPIRou’s exposure time calculator\footnote{\url{https://www.cfht.hawaii.edu/Instruments/SPIRou/SPIRou_etc.php}}, we estimate that detecting a $10^{-4}$ level polarization spectrum from Luhman 16A (binned to R = 100) requires 17.4 hours of integration under photon-noise–limited conditions.  In practice, systematic effects such as instrumental polarization, cross-talk, imperfect demodulation, and telluric residuals introduce additional noise floors. Achieving a robust $\sigma_p \simeq 10^{-4}$ therefore requires longer integration time, along with precise calibration and excellent instrumental stability. A detailed exposure time calculation is provided in Appendix B. Future instruments offering broadband, low-resolution linear spectropolarimetry would provide the most promising avenue for detecting polarization lower than $10^{-4}$ from faint brown dwarfs. By integrating over wide wavelength bands, such instruments could achieve the required polarimetric sensitivity within practical exposure time and directly probe the features illustrated in Figure ~\ref{fig14}.

\subsection{Insights into Atmospheric Characterization Through a Successful Spectropolarimetry Detection}\label{sec:6.3}

Detecting the disk-integrated polarization of a brown dwarf requires both strong intrinsic limb polarization and substantial atmospheric asymmetry across the disk. Features such as uneven cloud coverage, rotation-induced flattening, or lateral temperature inhomogeneities are essential for producing a detectable signal. Nonetheless, a successful detection of polarized thermal emission would be exceptionally informative, as it implies a specific combination of atmospheric parameters of cloud properties, temperature structure, and scattering geometry that collectively enhance the polarization signal.

In this study, we have examined the factors shaping disk-resolved polarized spectra. A detection across a broad wavelength range would capture the intricate interplay between temperature structure, cloud properties, and atmospheric composition. For spectropolarimetry to yield detectable signals in currently accessible wavelength bands (e.g., near-infrared), several atmospheric conditions must be satisfied.  
First, the cloud particle size must fall within the Rayleigh-like scattering regime, such that the size parameter \(x \sim 1\), ensuring efficient polarization from scattered thermal radiation. Second, the cloud optical thickness must be finely balanced, large enough to permit efficient single scattering, but not so high that multiple scattering causes significant depolarization. Third, the vertical temperature gradient in the photosphere must be sufficiently steep to produce the anisotropy required in the emergent thermal flux, thereby ruling out isothermal layers often inferred from flux-only retrievals. Together, these conditions are crucial for producing a polarization signal strong enough to be detected by current polarimetric instruments. Moreover, if a characteristic two-peak polarization signature is observed, it would provide direct constraints on the effective cloud particle radius and indicate that the cloud behaves as optically gray across the detected wavelength range.

A more straightforward approach to interpreting such polarization features would be to perform atmospheric retrievals directly on polarization spectra, analogous to flux-based retrievals. However, polarization retrievals are currently limited by the computational cost of polarization modeling. Polarization simulations are typically performed using three-dimensional Monte Carlo radiative transfer \citep{stolker2017polarized} or adding-doubling radiative transfer algorithm \citep{rossi2018pymiedap, chubb2024modelling} to take into account full orders of scattering. Monte Carlo simulations sample large numbers of photon packages (\(10^{9}\) in this paper) and track the polarization state of each emitted photon in a 3D spherical grid and estimate the final Stokes vector at the detector plane. This approach accurately captures the polarization state arising from 3D atmospheric asymmetries but is computationally expensive. While a 1D flux radiative transfer calculation may require only a fraction of a second to generate a near-infrared spectrum (1–2~$\mu$m), computing the polarization at each wavelength can take several minutes, depending on numbers of used photon packages and cloud opacities. Such computational costs render polarization retrievals, which typically require over a million iterations, currently infeasible. Several emerging methods have been developed to accelerate flux calculations, including GPU-optimized radiative transfer codes (e.g., \texttt{ExoJAX}, \texttt{gCMCRT}; \citealt{kawahara2022autodifferentiable,lee20223d}) and machine-learning–accelerated approaches (e.g., Neural posterior estimation; \citealt{vasist2023neural,martinez2024floppity,lueber2025near}). Applying these acceleration techniques to polarization calculations could enable sufficiently fast 3D polarization modeling for future retrieval frameworks, ultimately bridging the gap between theoretical polarization models and observational inference.

Although this study highlights that combining flux and polarization information can help constrain the atmospheric temperature structure and cloud scattering properties, interpreting the net polarization remains challenging. The observed polarization is a combined outcome of disk-resolved polarization (set by local atmospheric properties) and large-scale atmospheric asymmetries such as global cloud distributions. Three-dimensional general circulation models (GCMs) have demonstrated intricate couplings among temperature structure, cloud opacity, and chemical composition across the photosphere, giving rise to both latitudinal and temporal variability \citep{lee2023dynamically,lee2024dynamically}. These global asymmetries strongly modulate the net polarization through disk-averaging of spatially varying polarization signals. Unlike flux retrievals, where a simple cloud fraction can approximate the cloud coverage in observed hemisphere, polarization modeling requires a more sophisticated parameterization of the cloud distribution to correctly vector-average the linear polarization across the disk. Such parameterized asymmetries can either enhance or suppress the overall polarization signal and may introduce degeneracies between localized temperature structures and cloud properties that govern the disk-resolved polarization.

In summary, this study takes an incremental step toward the ambitious goal of fully retrieving atmospheric information from both intensity and polarization spectra. We clarify how the wavelength dependence of polarized thermal emission is governed by temperature profiles, molecular absorption, and cloud scattering properties. A schematic framework is proposed to show how these factors shape the observed spectrum.  Our key findings are as follows:

1. Broadband polarization features are primarily governed by the cloud size parameter and cloud optical thickness. Theoretically, variations in the cloud size parameter can produce a characteristic two-peak polarization signature, which serves as a first-order constraint on cloud particle size. For MgSiO$_3$ clouds, the first peak typically emerges near $x \sim 1$, and the second around $x \sim 0.2$. However, cloud optical thickness has a stronger influence and often dominates the broadband polarization trend.
Detecting both polarization peaks requires the cloud to behave as an optically gray medium across a broad spectral range, a condition generally satisfied only by clouds with large particle sizes. However, in such cases, the polarization peaks shift toward longer wavelengths, often beyond the near-infrared (e.g., $\geq 6.28\,\mu$m for $r_{\rm eff} \geq 1\,\mu$m). Depending on the slope of the cloud extinction cross-section with wavelength and the cloud mass density, only the peak associated with efficient single scattering without being diminished by excessive multiple scattering may appear prominently in the polarization spectrum.

2. Narrow-band peaks and troughs polarization features are shaped by molecular opacity and the temperature structure. These narrower features in molecular absorption band in principle trace the temperature gradient at the photosphere at each wavelength. Even if different temperature profiles and molecular parameter combinations may yield similar flux spectra, their polarization signatures can differ significantly, especially within absorption bands. This narrow-band polarization features offer a unique tool to possibly reduce the degeneracies between temperature structure and gas abundances commonly encountered in flux spectral retrievals.

These findings reinforce the diagnostic power of spectropolarimetry in probing atmospheric structure, particularly when used alongside flux measurements. As next-generation extremely large telescopes (ELTs) like the ESO ELT, TMT, and GMT come online, their increased collecting areas and improved polarimetric sensitivity (down to \(\sim 10^{-6}\)) will make it possible to detect the faint polarized emission from brown dwarfs and self-luminous exoplanets. Instruments such as EPOL \citep{keller2010epol}, proposed for the ELT, promise to  have an integral field unit to obtain linearly polarized spectra between 600 and 900 nm. Future instruments offering broadband, low-resolution linear spectropolarimetry would provide the most promising avenue for detecting polarization features predicted in this paper. These capabilities will pave the way that incorporating both the scalar and polarized components of thermal radiation into of atmospheric retrievals, unlocking detailed insights into particle size distributions, vertical temperature structures of the atmosphere.

\section*{Acknowledgments}

FW and BB acknowledge support from UK Research and Innovation Science and Technology Facilities Council [ST/X001091/1].

\section*{Data Availability}

All data underlying this article are publicly available upon reasonable request to the corresponding author.


\bibliographystyle{mnras}
\bibliography{ref} 

=



\appendix

\section{Retrieval Setup and Result }

\begin{figure*}
	\includegraphics[width=2.1\columnwidth]{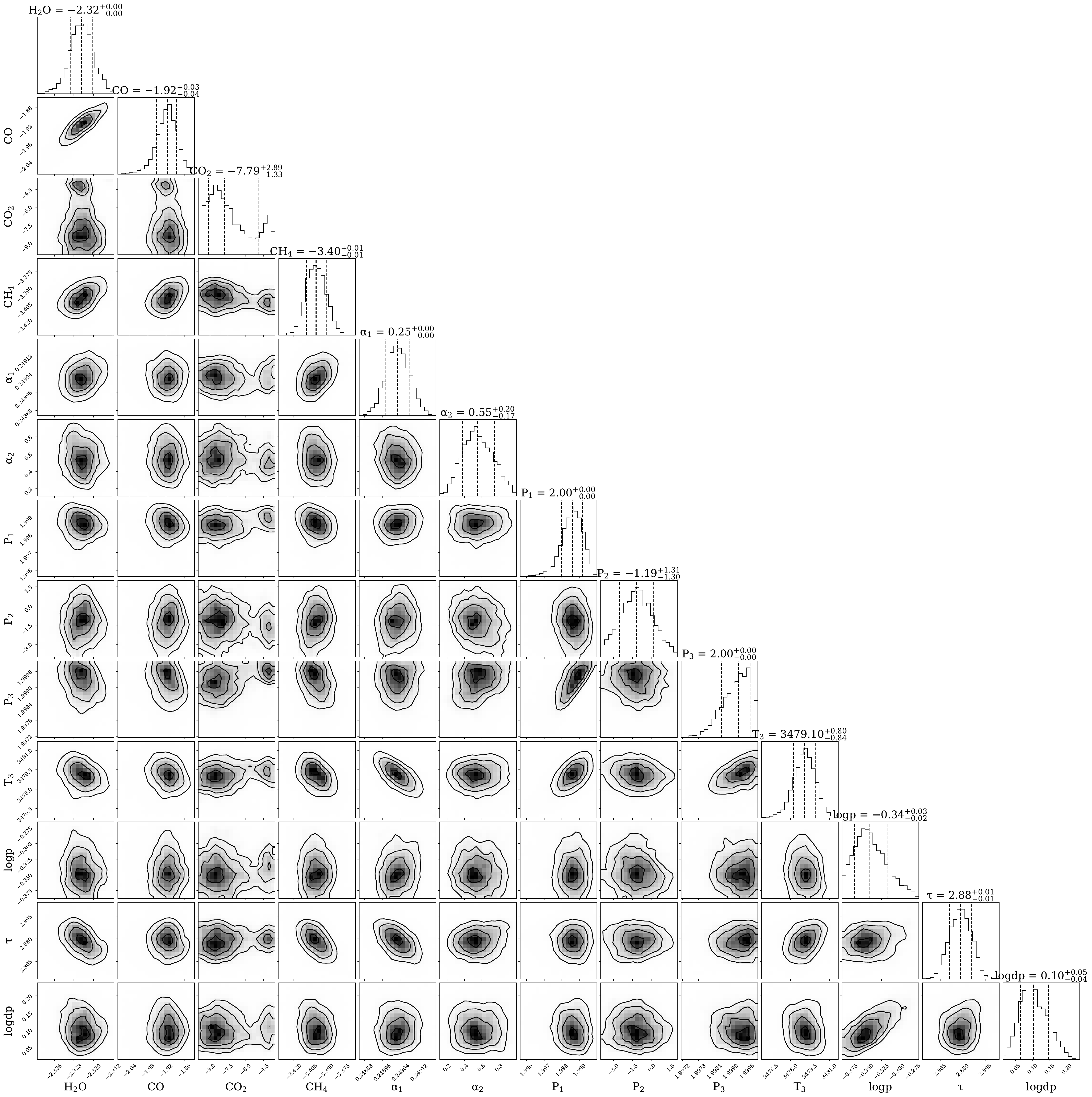}
    \caption{Marginalized posterior probability distributions for the degenerated gas abundances and $P\text{--}T$ profile parameters. The values above the 1–D histograms represents the parametric median (50th percentile) values with the errors representing the 1$\rm \sigma$ central credible interval (16th and 84th percentile) values. The different shades in the 1–D and 2–D histograms represent the 1, 2 and 3$\rm \sigma$ central credible interval, respectively, with the darkest shade corresponding to 1$\rm \sigma$.}
    \label{fig:corner}
\end{figure*}

\begin{figure*}
	\includegraphics[width=1.5\columnwidth]{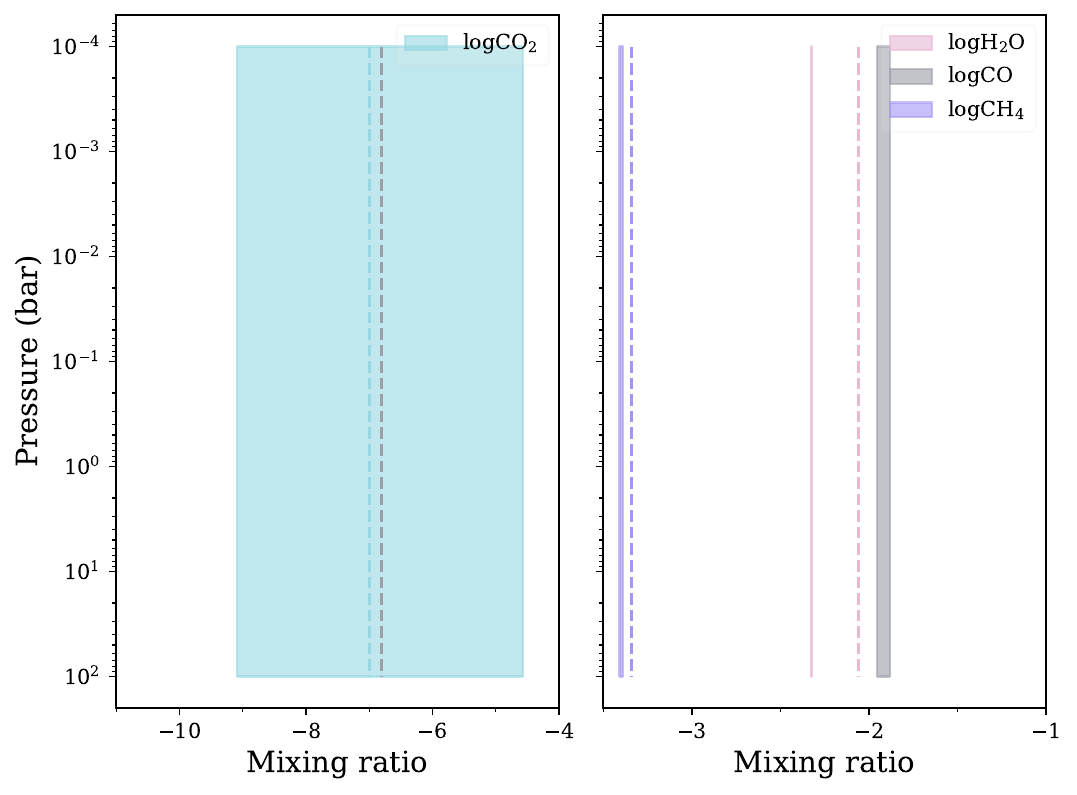}
    \caption{Comparison of the 1$\sigma$ retrieved abundances (shaded) with the input values for H$_2$O, CO, CO$_2$, and CH$_4$. The dominant gases, H$_2$O and CH$_4$, are more abundant in the input case, whereas CO is five orders of magnitude less abundant in the degenerate scenario.}
    \label{fig:abun}
\end{figure*}

Table~\ref{tab:priors} lists the prior ranges for all retrieved model parameters. Figure~\ref{fig:corner} shows marginalized posterior probability distributions all parameters, and the difference of gas abundances between the input and retrieved degenerated solutions are summarized in 
Figure~\ref{fig:abun}.

\begin{table*}
\centering
\begin{threeparttable}
\caption{Priors used for the retrieval setup in Section~\ref{sec:5.3}}
\label{tab:priors}
\begin{tabular}{ll}
\toprule\toprule
\textbf{Parameter} & \textbf{Prior\tnote{a}} \\
\midrule

Gas Volume Mixing Ratio \text{--} $\log(f_i)$\tnote{b,c} 
& $\mathcal{U}(-12, 0),\quad \sum_{i=1}^{4} f_i \le 1$ \\

Madhusudhan \& Seager Thermal Profile: 
$\alpha_1, \alpha_2, P_1, P_2, P_3, T_3$ 
& $\mathcal{U}(0.25, 0.5),\ \mathcal{U}(0.1, 0.2),\ \mathcal{U}(10^{-4}, 10^{2.3}),\ \mathcal{U}(10^{-4}, 10^{2.3}),\ \mathcal{U}(10^{-4}, 10^{2.3}),\ \mathcal{U}(0, 5000)$ \\

Cloud top pressure \text{--} $\log p$ 
& $\mathcal{U}(-4, 2.3)$ \\

Cloud optical thickness \text{--} $\tau$ 
& $\mathcal{U}(0, 10)$ \\

Cloud thickness \text{--} $\log dp$ 
& $\mathcal{U}(0, 1)$ \\

\bottomrule
\end{tabular}

\begin{tablenotes}
\footnotesize
\item[a] $\mathcal{U}$ denotes a uniform prior.
\item[b] The gas species considered.
\item[c] The mixing ratios are constrained such that $\sum f_i \le 1$.
\end{tablenotes}

\end{threeparttable}
\end{table*}

\section{Exposure time estimate on a $10^{-4}$ level H band polarization  with SPIRou}

Luhman~16 (WISE~J104915.57−531906.1AB; \citealt{luhman2013discovery}) is the closest known brown dwarf system to Earth, located at a distance of $\sim$1.99 pc. The binary is of particular interest because its two components span the L/T transition, with spectral types of L7.5$\pm$1 and T0.5$\pm$1 for components A and B, respectively. \citet{millar2020detection} reported H-band linear polarization detections for both components, with with \(P_{\rm A} = 0.031\%\) and \(P_{\rm B} = 0.01\%\). Motivated by this $\sim10^{-4}$ level polarization, we assess the feasibility of performing near-infrared spectropolarimetric observations of Luhman~16A with SPIRou.

Polarimetry relies on detecting small fractional differences between intensities measured in orthogonal polarization states. For a target with an H-band polarization of $\sim10^{-4}$, achieving a fractional polarization sensitivity of this level requires a signal-to-noise ratio (S/N) of at least 10000. SPIRou is a near-infrared (NIR) spectropolarimeter on Canada–France–Hawaii Telescope (CFHT). It provides high-resolution ($R \sim 70000$) spectropolarimetry over the wavelength range 0.98–2.35~$\mu$m. In practice, high-resolution spectropolarimetry of faint or cool brown dwarfs with SPIRou is strongly S/N-limited: achieving $\mathrm{S/N} \approx 10000$ per native-resolution pixel is not feasible for typical targets with $J \gtrsim 13$. Low-resolution ($R \lesssim 100$) polarimetry would be a more practical approach.

\subsection{Spectra binning}
By integrating over SPIRou’s high-resolution spectral elements, we estimate the exposure times required for low-resolution ($R=100$) spectropolarimetric observations capable of reaching a polarization sensitivity of $10^{-4}$.
The SPIRou Exposure Time Calculator (ETC) uses synthetic spectra from the PHOENIX library (2300~K~$<~T_{\rm eff}~<~8000$~K) combined with atmospheric transmission templates to estimate the exposure time needed for a given S/N per 2.28~km/s bin. As a reference, we adopt a star with $H = 9.56$~mag and $T_{\rm eff} = 2300$~K, for which the ETC yields a $\mathrm{S/N}(H) = 90$ for an exposure time of $t_0 = 1777.4$~s per spectral bin at $1.65~\mu$m (center of the $H$ band).

Luhman~16A has a same H-band magnitude but a cooler temperature ($T_{\rm eff} = 1320$~K). Scaling from the ETC reference, we estimate the exposure time required to achieve $\sigma_p \approx 10^{-4}$. Starting from $\mathrm{S/N}_{\rm pix} = 90$ per $2.28~\mathrm{km/s}$ pixel in $t_0 = 1777.4$~s, coarsening to a low-resolution mode ($R = 100$) corresponds to combining $N \approx 700$ native resolution elements. The photon-limited S/N then becomes

\begin{equation}
\mathrm{S/N}_{R100,\,\mathrm{raw}} = \mathrm{S/N}_{\rm pix} \times \sqrt{N} \approx 90 \times \sqrt{700} \approx 2380
\end{equation}

Accounting for the four-exposure polarimetric modulation, we include a conservative degradation factor of $\sqrt{2}$:

\begin{equation}
\mathrm{S/N}_{R100} = \frac{\mathrm{S/N}_{R100,\,\mathrm{raw}}}{\sqrt{2}} \approx 1680
\end{equation}

Thus, for $t_0 = 1777.4$~s, we obtain $\mathrm{S/N}_{R100} \approx 1.7\times10^3$. To reach $\sigma_p = 10^{-4}$ requires at least $\mathrm{S/N}_{\rm req}=1/\sigma_p=10000$, the required exposure time is

\begin{equation}
t = t_0 \left( \frac{\mathrm{S/N}_{\rm req}}{\mathrm{S/N}_{R100}} \right)^2
   = 1777.4 \times \left( \frac{10000}{1683.75} \right)^2
   \approx 17.4~\mathrm{h}
\end{equation}

This $\sim$17 hour estimate represents the ideal photon-noise limit. In practice, systematic effects such as instrumental polarization, cross-talk, imperfect demodulation, and telluric residuals introduce additional noise floors. Achieving a robust $\sigma_p \simeq 10^{-4}$ therefore demands longer integrations and highly stable, well-calibrated instrumentation. Furthermore, Luhman~16 exhibits multi-hour rotational variability due to patchy cloud structures. Integrating over 17 hours would smear out rotational phase-dependent polarization signals. For phase-resolved polarimetry, the total integration should thus be divided into shorter observing segments, accepting correspondingly higher per-segment noise.



\bsp	
\label{lastpage}
\end{document}